\documentclass[12pt]{article}
\usepackage{latexsym}
\begin{document}
      \sloppy

\def\AFOUR{%
\setlength{\textheight}{9.0in}
\setlength{\textwidth}{5.75in}%
\setlength{\topmargin}{-0.375in}%
\hoffset=-.5in%
\renewcommand{\baselinestretch}{1.17}%
\setlength{\parskip}{6pt plus 2pt}%
}
\AFOUR
\def\car{\mathop{\square}}
\def\carre#1#2{\raise 2pt\hbox{$\scriptstyle #1$}\car_{#2}}

\parindent=0pt
\makeatletter
\def\section{\@startsection {section}{1}{\z@}{-3.5ex plus -1ex minus
   -.2ex}{2.3ex plus .2ex}{\large\bf}}
\def\subsection{\@startsection{subsection}{2}{\z@}{-3.25ex plus -1ex minus
   -.2ex}{1.5ex plus .2ex}{\normalsize\bf}}
\makeatother
\makeatletter
\@addtoreset{equation}{section}
\renewcommand{\theequation}{\thesection.\arabic{equation}}
\makeatother

\renewcommand{\a}{\alpha}
\renewcommand{\b}{\beta}
\newcommand{\g}{\gamma}           \newcommand{\G}{\Gamma}
\renewcommand{\d}{\delta}         \newcommand{\D}{\Delta}
\newcommand{\e}{\varepsilon}
\newcommand{\la}{\lambda}        \newcommand{\LA}{\Lambda}
\newcommand{\m}{\mu}
\newcommand{\A}{\widehat{A}^{\star a}_{\mu}}
\newcommand{\Ar}{\widehat{A}^{\star a}_{\rho}}
\newcommand{\n}{\nu}
\newcommand{\om}{\omega}         \newcommand{\OM}{\Omega}
\newcommand{\p}{\psi}             \newcommand{\PS}{\Psi}
\renewcommand{\r}{\rho}
\newcommand{\s}{\sigma}           \renewcommand{\S}{\Sigma}
\newcommand{\vf}{\varphi}
\newcommand{\x}{\xi}              \newcommand{\X}{\Xi}
\renewcommand{\x}{\xi}              \renewcommand{\X}{\Xi}
\newcommand{\y}{\upsilon}       \newcommand{\Y}{\Upsilon}
\newcommand{\z}{\zeta}

\renewcommand{\AA}{{\cal A}}
\newcommand{\BB}{{\cal B}}
\newcommand{\CC}{{\cal C}}
\newcommand{\DD}{{\cal D}}
\newcommand{\EE}{{\cal E}}
\newcommand{\FF}{{\cal F}}
\newcommand{\GG}{{\cal G}}
\newcommand{\HH}{{\cal H}}
\newcommand{\II}{{\cal I}}
\newcommand{\JJ}{{\cal J}}
\newcommand{\KK}{{\cal K}}
\newcommand{\LL}{{\cal L}}
\newcommand{\MM}{{\cal M}}
\newcommand{\NN}{{\cal N}}
\newcommand{\OO}{{\cal O}}
\newcommand{\PP}{{\cal P}}
\newcommand{\QQ}{{\cal Q}}
\renewcommand{\SS}{{\cal S}}
\newcommand{\RR}{{\cal R}}
\newcommand{\TT}{{\cal T}}
\newcommand{\UU}{{\cal U}}
\newcommand{\VV}{{\cal V}}
\newcommand{\WW}{{\cal W}}
\newcommand{\XX}{{\cal X}}
\newcommand{\YY}{{\cal Y}}
\newcommand{\ZZ}{{\cal Z}}

\newcommand{\wh}[1]{\widehat{#1}}
\newcommand{\fd}[1]{\frac{\d}{\d #1}}
\newcommand{\ch}{\widehat{C}}
\newcommand{\gh}{\widehat{\gamma}}
\newcommand{\W}{W_{i}}
\newcommand{\na}{\nabla}
\newcommand{\xint}{\dint d^4x\;}
\newcommand{\sla}{\raise.15ex\hbox{$/$}\kern -.57em}
\newcommand{\Sla}{\raise.15ex\hbox{$/$}\kern -.70em}
\def\h{\hbar}
\def\Lp{\displaystyle{\biggl(}}
\def\Rp{\displaystyle{\biggr)}}
\def\LP{\displaystyle{\Biggl(}}
\def\RP{\displaystyle{\Biggr)}}
\newcommand{\lp}{\left(}\newcommand{\rp}{\right)}
\newcommand{\lc}{\left[}\newcommand{\rc}{\right]}
\newcommand{\lac}{\left\{}\newcommand{\rac}{\right\}}
\newcommand{\identity}{\bf 1\hspace{-0.4em}1}
\newcommand{\complex}{{\kern .1em {\raise .47ex
\hbox {$\scriptscriptstyle |$}}
      \kern -.4em {\rm C}}}
\newcommand{\real}{{{\rm I} \kern -.19em {\rm R}}}
\newcommand{\rational}{{\kern .1em {\raise .47ex
\hbox{$\scripscriptstyle |$}}
      \kern -.35em {\rm Q}}}
\renewcommand{\natural}{{\vrule height 1.6ex width
.05em depth 0ex \kern -.35em {\rm N}}}
\newcommand{\tint}{\int d^4 \! x \, }
\newcommand{\intg}{\int d^D \! x \, }
\newcommand{\intm}{\int_\MM}
\newcommand{\tr}{{\rm {Tr} \,}}
\newcommand{\half}{\dfrac{1}{2}}
\newcommand{\f}{\frac}
\newcommand{\pa}{\partial}
\newcommand{\pad}[2]{{\frac{\partial #1}{\partial #2}}}
\newcommand{\fud}[2]{{\frac{\delta #1}{\delta #2}}}
\newcommand{\dpad}[2]{{\displaystyle{\frac{\partial #1}{\partial
#2}}}}
\newcommand{\dfud}[2]{{\displaystyle{\frac{\delta #1}{\delta #2}}}}
\newcommand{\dfrac}[2]{{\displaystyle{\frac{#1}{#2}}}}
\newcommand{\dsum}[2]{\displaystyle{\sum_{#1}^{#2}}}
\newcommand{\dint}{\displaystyle{\int}}
\newcommand{\eg}{{\em e.g.,\ }}
\newcommand{\Eg}{{\em E.g.,\ }}
\newcommand{\ie}{{{\em i.e.},\ }}
\newcommand{\Ie}{{\em I.e.,\ }}
\newcommand{\nb}{\noindent{\bf N.B.}\ }
\newcommand{\etal}{{\em et al.}}
\newcommand{\etc}{{\em etc.\ }}
\newcommand{\via}{{\em via\ }}
\newcommand{\cf}{{\em cf.\ }}
\newcommand{\twiddle}{\lower.9ex\rlap{$\kern -.1em\scriptstyle\sim$}}
\newcommand{\qed}{\vrule height 1.2ex width 0.5em}
\newcommand{\grad}{\nabla}
\newcommand{\bra}[1]{\left\langle {#1}\right|}
\newcommand{\ket}[1]{\left| {#1}\right\rangle}
\newcommand{\vev}[1]{\left\langle {#1}\right\rangle}
\newcommand{\wt}[1]{\widetilde{#1}}

\newcommand{\equ}[1]{(\ref{#1})}
\newcommand{\be}{\begin{equation}}
\newcommand{\ee}{\end{equation}}
\newcommand{\eqn}[1]{\label{#1}\end{equation}}
\newcommand{\eea}{\end{eqnarray}}
\newcommand{\bea}{\begin{eqnarray}}
\newcommand{\eqan}[1]{\label{#1}\end{eqnarray}}
\newcommand{\ba}{\begin{array}}
\newcommand{\ea}{\end{array}}
\newcommand{\eqac}{\begin{equation}\lac\begin{array}{rcl}}
\newcommand{\eqacn}[1]{\end{array}\right.\label{#1}\end{equation}}
\newcommand{\qq}{&\qquad &}
\renewcommand{\=}{&=&} 
\newcommand{\cb}{{\bar c}}
\newcommand{\mn}{{\m\n}}
\newcommand{\pic}{$\spadesuit\spadesuit$}
\newcommand{\?}{{\bf ???}}
\newcommand{\Tr }{\mbox{Tr}\ }
\newcommand{\adot}{{\dot\alpha}}
\newcommand{\bdot}{{\dot\beta}}
\newcommand{\gdot}{{\dot\gamma}}

\newcommand{\df}[1]{\lp\pa_3\vf\rp_{#1}}
\newcommand{\dft}[1]{\lp \wt{\pa_3\vf}\rp_{#1}}


\def\bra#1{\left\langle #1\right|}
\def\ket#1{\left| #1\right\rangle}

\def\ha{\frac{1}{2}}

\def\ov{\overline}

\def\und{\underline}

\def\ve{\varepsilon}

\def\a{\alpha}
\def\b{\beta}
\def\g{\gamma}
\def\G{\Gamma}
\def\e{\epsilon}
\def\f{\phi}
\def\fb{{\ov \phi}}
\def\vf{\varphi}
\def\m{\mu}
\def\n{\nu}
\def\r{\rho}
\def\k{\kappa}
\def\dk{\dot\kappa}
\def\s{\sigma}
\def\sb{\ov\sigma}
\def\l{\lambda}
\def\L{\Lambda}
\def\p{\psi}
\def\pb{\ov\psi}
\def\cb{\ov\chi}
\def\d{\partial}
\def\dag{\dagger}
\def\dalpha{{\dot\alpha}}
\def\dbeta{{\dot\beta}}
\def\dgamma{{\dot\gamma}}
\def\ddelta{{\dot\delta}}
\def\da{{\dot\alpha}}
\def\db{{\dot\beta}}
\def\dg{{\dot\gamma}}
\def\dd{{\dot\delta}}
\def\t{\theta}
\def\tb{{\ov \theta}}
\def\lb{{\ov \lambda}}
\def\eb{{\ov \epsilon}}
\def\zb{{\ov z}}
\def\ib{{\ov i}}
\def\jb{{\ov j}}
\def\kb{{\ov k}}
\def\mb{{\ov m}}
\def\nb{{\ov n}}
\def\qb{{\ov q}}
\def\xib{\ov \xi}
\def\tq{\tilde{q}}
\def\D{\Delta}
\def\DD{\Delta^\dag}
\def\Db{\ov D}
\def\M{{\cal M}}
\def\rd{\sqrt{2}}
\def\Tr{{\rm Tr\, }}
\def\F{{\cal F}}
\def\Dint{\int\, {\rm d}^2\theta {\rm d}^2\overline\theta\ }
\def\Fint{\int\, {\rm d}^2\theta\ }
\def\Fbarint{\int\, {\rm d}^2\overline\theta\ }
\def\xint{\int\, {\rm d}^4 x}
\def\Dcomp{{\theta\theta\ov{\theta\theta}}}
\def\Fcomp{{\theta\theta}}
\def\Fbarcomp{{\ov{\theta\theta}}}
\def\dw{\delta_W}
\def\dmu{\delta_\mu}
\def\dnu{\delta_\nu}
\def\dmn{\delta_{\mu\nu}}
\def\drs{\delta_{\rho\sigma}}
\def\H{{\rm H}}
\def\ub{\ov u}
\def\vb{\ov v}
\def\Hb{\ov \H}
\def\app{\ \epsilon\ }

\global\parskip=4pt
\titlepage  \noindent
{
   \noindent

\hfill GEF-TH-4/2010


\vspace{1cm}

\noindent
{\bf
{\large Let's Twist Again: N=2 Super Yang-Mills Theory 

Coupled To Matter
}}

\vspace{.5cm}
\hrule

\vspace{1cm}

\noindent

\begin{center}{\bf 
Nicola Maggiore}
\footnote{nicola.maggiore@ge.infn.it}\end{center}
\begin{center}{\footnotesize {\it
 Dipartimento di Fisica, Universit\`a di Genova \\
via Dodecaneso 33, I-16146 Genova -- Italy \\and \\INFN, Sezione di
Genova 
} }
\end{center}

\begin{center}{\bf Marco Picollo }
\footnote{mpicollo@selex-si.com}\end{center}
\begin{center}{\footnotesize {\it
 Dipartimento di Fisica, Universit\`a di Genova \\
via Dodecaneso 33, I-16146 Genova -- Italy 
} }
\end{center}

\noindent
{\tt Abstract~:}
We give the twisted version of N=2 Super Yang-Mills theory coupled to matter, 
including quantum fields, supersymmetry transformations, action and algebraic 
structure. We show that the whole action, coupled to matter, can be written as 
the variation of a nilpotent operator, $modulo$ field equations. 
An extended Slavnov-Taylor identity , collecting gauge symmetry and 
supersymmetry, is written, which allows to define the web of algebraic 
constraints, in view of the algebraic renormalization and of the extension of the algebraic proof 
of the non-renormalization theorems holding for N=2 SYM theory without matter.

\vfill\noindent
{\footnotesize {\tt Keywords:}
Topological Quantum Field Theories,
Extended Supersymmetry,
BRST Renormalization.
\\
{\tt PACS Nos:} 
 03.70.+k (theory of quantized fields), 
11.10.Gh	(renormalization),\\
 11.30.Pb (supersymmetry),
 12.60.Jv (supersymmetric models).
}
\newpage
\begin{small}
\end{small}

\setcounter{footnote}{0}


\section{Introduction}

Before 1994, the reasons to discard N=2 Super Yang-Mills (SYM) theory prevailed 
over those for studying it. This is testified by the weak 
occurrence in the literature before then, of theories 
with extended supersymmetries.  

Concerning phenomenology, the presence of the so called ``mirror particles'' 
eliminates every 
possible physical interest: fermions of opposite chirality,
{\it but in the same representation} of the gauge group, unavoidably appear in 
the theory, 
which hence is not chiral, and consequently not realistic, if one wants to include 
the particles of the Standard Model \cite{West:1990tg}.

From the Quantum Field Theory (QFT) point of view, 
on the other hand, theories with extended 
supersymmetry 
represent a real challenge, as explained in \cite{Breitenlohner:1985kh}.

In fact, while  for N=1 SYM the superspace formalism based on unconstrained superfields allowed to perform the algebraic quantum extension of the theory \cite{Piguet:1986ug}, the superfield approach to theories with extended supersymmetry is troublesome for several reasons. N=2 supersymmetry can be realized by means of of N=1 superfields, but the necessary additional symmetry involving N=1 superfields in non-polynomial. On the other hand, the harmonic superspace approach \cite{Galperin:2001uw} is possible, but a regularization scheme preserving both supersymmetry and gauge invariance, to all orders of perturbation theory, is still lacking. Despite this,  the most celebrated results concerning the good renormalization properties of theories with extended supersymmetry, in particular the vanishing of the $\beta$-function above one loop, have been obtained in a superspace (N=1 and/or N=2) framework \cite{Grisaru:1982zh,Howe:1983sr,Howe:1983wj}. A review of these results, of the ways employed to get them and also of as the weaknesses of each of them, can be found in Chapter 18 of \cite{West:1990tg}\footnote{ At pag 194 of \cite{West:1990tg}, it is pointed out: ``Here we have stressed these weaknesses not because of a mistrust in the arguments for 
finiteness, but to show that they are not proofs in a mathematical sense and that there is 
still room for further work''. }.

The situation doesn't sound much better in components. The drawback of adopting the  WZ 
gauge, is that
the supersymmetry transformations are nonlinear, and the 
supersymmetry algebra does not  close on translations, but two kinds of
obstructions occur: field dependent gauge transformations and field 
equations of motion. This fact has two consequences: the difficulty 
of defining a gauge fixing term, which is invariant under both 
supersymmetry and BRS symmetry, and the need of an infinite 
number of external sources, with increasing negative mass dimensions, in order to 
control the algebra \cite{Breitenlohner:1985kh}.

After the appearance of  the celebrated Seiberg-Witten papers 
\cite{Seiberg:1994rs, Seiberg:1994aj} on the 
{\it electric-magnetic duality} in N=2 SYM theory, which relates the 
weak and strong coupling regimes of that theory, the N=2 susy 
theories faced a kind of second youth, becoming extremely popular, and 
were massively reconsidered by the community.

Most of the problems described in \cite{Breitenlohner:1985kh} 
were solved, and the renormalizability 
of N=2 coupled to matter, by means of a non-anomalous Slavnov-Taylor identity, was rigorously established 
\cite{Maggiore:1994dw, Maggiore:1994xw}, using a 
technique which has been successfully repeated since, 
and which we are adopting also in this paper 
for the classical definition of the theory (see Section 5).

More recently, the method of ``shadow fields'' has been introduced, which allowed to write a system of Slavnov-Taylor identities by means of which supersymmetric gauge field theories can be renormalized in a regularization independent way, permitting also to study the observables which are not scalar under supersymmetric transformations \cite{Baulieu:2006gx}.

Other important goals have been reached exploiting the $twist$ 
\cite{Witten:1988ze,Witten:1994ev}. 
Indeed, it was a known fact  
that N=2 SYM is related to topological field theories, in particular 
Topological Yang-Mills (TYM) theory, by means of a twist, 
which ultimately reduces to a linear redefinition of the 
quantum fields, to which the path integral defining the generating 
functionals is insensitive (we shall be more precise in 
Section 3). Consequently, the twisted-related 
theories are completely equivalent. 

The bad news driven by this, is 
that the results concerning N=2,4 SYM theories can hardly be extended to 
the more realistic N=1 SYM theories, which are definitely not 
topological QFTs, having local degrees of freedom. 

This drawback is partially 
compensated by the fact that some important facts concerning N=2 
SYM can be proved through their equivalent twisted version: TYM. This has been 
the case, for instance, for the theorem concerning the N=2 SYM $\beta$-function, whose 
finiteness above one loop has been algebraically demonstrated in \cite{Blasi:2000qw} exploiting the existence of the twist.  

Later, it has also been algebraically proved by means of the shadow technique that, chosen the matter hypermultiplet in order to have a vanishing $\beta$-function at one loop, it vanishes at all orders of perturbation theory \cite{Baulieu:2007dk}.

Moreover, a central 
role is played by the operator $\Tr\phi^{2}$, which, in the 
Seiberg-Witten supersymmetric theory, is the gauge invariant quantity parametrizing 
the space of vacua of the theory, and, in the twisted topological theory, is the 
finite operator \cite{Maggiore:2001zw} by means of which the pure gauge theory can be 
defined \cite{Blasi:2000qw,Fucito:1997xm}.

So far the state of the art on which this paper stands. The program 
is not yet completely carried out: the non-renormalization theorems 
concerning the $\beta$-function are algebraically proved for the N=2 SYM, {\it in 
absence of matter}.  We recall that matter is coupled to the pure gauge 
theory by means of the hypermultiplet \cite{West:1990tg}, in a generic 
representation of the gauge group. As a consequence of the second 
supersymmetry, the theory, even in presence of matter, has 
$only\ one$ coupling constant. It is 
natural to ask which is the fate of the non-renormalization theorem 
concerning the unique $\beta$-function
in presence of matter. Related to this, it is interesting to know if, 
as in the pure gauge case, the whole theory can be written in terms of a 
single operator, which is finite to all orders of perturbation theory, 
and, if the answer is positive, which 
this operator is. Finally, the inclusion of matter allows 
also for taking into account N=4 SYM, which can be reached from N=2 
in the particular case of matter in the adjoint, rather than 
generic, representation of the gauge group. 

The aim of this paper, is to contribute to answer these questions. 
The preliminary and necessary step is to give the complete twisted version of 
N=2 SYM {\it coupled to matter}, and to achieve the whole set up for its 
quantum extension (gauge fixing, BRS symmetry, Slavnov-Taylor 
identity, algebraic structure, etc.) \cite{Piguet:1995er}. 

The paper is organized as follows. In Section 2 we recall the basics 
of N=2 SYM theories, with and without matter. In Section 3 we 
introduce the twist for the pure gauge case. The main results of this 
paper are contained in Sections 4 and 5, where the twisted version of 
the whole theory, including matter, is given, as well as the basis 
for the quantum implementation, which relies on the extended Slavnov-Taylor 
identity and on the {\it off shell} closed algebra. Conclusions and 
perspectives are summarized in  \mbox{Section 6.}

\section{The untwisted theory: N=2 SYM coupled to matter}

\subsection{Pure N=2 SYM}

The N=2 susy algebra reads
\bea
\{\QQ^{i}_{\a},\overline{\QQ}_{j\adot}\}&=&                                      
\delta^{i}_{j}(\s^{\m})_{\a\adot}\partial_{\m} 
\nonumber \\
\{\QQ^{i}_{\a},\QQ^{j}_{\b}\} &=& 
                   \{\overline{\QQ}_{i\adot},\overline{\QQ}_{j\bdot}\} = 0\ ,
		   \label{n=2algebra}
\eea
where $(\QQ^{i}_{\a},\overline{\QQ}_{j\adot})$ are the supersymmetry 
charges, 
indexed by $i=1,2$ and Weyl spinor indices $\a,\adot = 1,2$. The 
total 
number of supercharges is therefore eight.

The pure N=2 SYM theory is based on the Yang-Mills (YM) 
multiplet~\cite{West:1990tg}, which belongs to 
the adjoint representation of the gauge group, and  
whose field components are 
$(A_{\m},\lambda^{i\a},\overline{\lambda}_{i\adot},\phi,\overline\phi)$, 
where $A_{\m}(x)$ is the gauge field, 
$\lambda^{i\a}(x),\overline{\lambda}_{i\adot}(x)$ are two pairs 
of Weyl spinors, and 
$\phi(x),\overline\phi(x)$ are two scalars. 

The corresponding pure N=2 SYM action reads
\begin{eqnarray} \label{SYMN2pg}
S_{YM}&=&\frac{1}{g^2} \tr\xint \Big( \half
F_{\m\n}F^{\m\n}-4\l^{i\a}\s^\m_{\a\da}D_\m\lb_i^\da 
- \ha\fb D_\m D^\m \f 
\nonumber \\
&& - \fb \left\{ \lb^{i\da},\lb_{i\da} \right\}
+\f \left\{ \l^{i\a},\l_{i\a} \right\} - \frac{1}{32} \left[
\f,\fb \right] \left[ \f,\fb \right] \Big)\ ,\label{sym}
\end{eqnarray} 
where the Trace $\tr$ is done over the adjoint representation group.

The global symmetry group of the theory is 
\begin{equation}
\label{globalgroup}
H=SU(2)_L\times SU(2)_R\times SU(2)_I\times U(1)\ ,
\label{H}
\end{equation}
where $SU(2)_L\times SU(2)_R$ represents the Lorentz group, 
$SU(2)_I\times U(1)$ is the internal symmetry group,
$SU(2)_I$ referring to the supersymmetry index $i=1,2$ and $U(1)$ 
being the rigid ${\cal R}$--symmetry.

Correspondingly, the fields belonging to the N=2 YM multiplet
are assigned the following $H$--group quantum numbers:
\begin{eqnarray}\label{trasfunderH}
    A_\m &:& \left(\ha,\ha,0\right)^0 \nonumber\\
\l^{i\a} &:&\left(\ha,0,\ha\right)^{-1} \nonumber\\
\lb_{i\da} &:& \left(0,\ha,\ha \right)^{+1} \\
\f &:& \left(0,0,0\right)^{+2} \nonumber\\
\fb &:& \left(0,0,0\right)^{-2}\nonumber\ ,
\end{eqnarray}
where we adopted the notation 
\bea
\left (
SU(2)_L , SU(2)_R, SU(2)_I
\right ) ^{U(1)}\ .
\label{qmnotation}\eea

For what concerns the supersymmetry generators, the quantum numbers are:

\begin{equation}\label{rapprHQ}
Q_{i\a}=\left(\ha,0,\ha \right)^{+1} \quad ; \quad \ov
Q_{i\da}=\left(0,\ha,\ha \right)^{-1}\ .
\end{equation}

The supersymmetry transformations of the pure N=2 SYM fields are:
\begin{eqnarray}\label{susytransfN2pg}
\delta {A}_\m&=&-\rd\xi^{\a j}(\s_\m)_{\a\da}\lb_j^\da -\rd
\xib^{\da j}(\s_\m)_{\a\da} \l_i^\a \nonumber\\  
\delta\f&=&-4\rd\xib^{\da j}\lb_{j \da} \nonumber \\
\delta\fb&=&-4\rd\xi^{\a j}\l_{j \a} \\
\delta\l_{i \k}&=&\frac{\rd}{8}\xi^{\a
j}\ve_{\a\k}\ve_{ij}\left[\f,\fb\right]+\frac{1}{\rd}\xi^{\a
j}\ve_{ij}(\s^{\m\n})_{\k\a}{F}_{\m\n}+\frac{1}{\rd}\xib^{\da
j}\ve_{ij}(\s_\m)_{\k\da}{D}^\m\fb \nonumber \\
\delta\lb_{i \dk}&=&\frac{1}{\rd}\xi^{\a
j}\ve_{ij}(\s_\m)_{\a\dk}{D}^\m\f+\frac{\rd}{8}\xib^{\da
j}\ve_{\dk\da}\ve_{ji}\left[\f,\fb\right]+\frac{1}{\rd}\xib^{\da
j}\ve_{ji}(\sb^{\m\n})_{\dk\da}{F}_{\m\n}\ , \nonumber
\end{eqnarray}
where the operator $\delta$ collects the supercharges $Q_{i\a}$ and 
$Q_{i \da} $ through \\
$\delta=\xi^{\a j} Q_{j\a}+\xib^{\da j} \ov Q_{j \da}$.
Notice that in the Wess-Zumino gauge the supersymmetry 
transformations \equ{susytransfN2pg} are nonlinear.

The action (\ref{SYMN2pg}) is susy invariant :
\bea
\delta {S}_{YM}=0\ .
\eea

\subsection{N=2 SYM coupled to matter}

To couple pure N=2 SYM to matter, we need the matter hypermultiplet
$(q_{i},\tq_{i},\psi_{q},\pb_{q},\p_{\tq},\pb_{\tq})$ \cite{West:1990tg}, 
formed by two pairs of scalar fields $q_{i}(x)$
and $\tq_{i}(x)$, two Weyl fermions $\psi_q(x)$ and $\psi_{\tq}(x)$ and their 
complex conjugates, all in a generic complex representation of the gauge 
group. The matter $H$--quantum numbers \equ{qmnotation} are:  
\begin{eqnarray}\label{transunderH2}
q_i &:& (0,0,\ha)^0 \nonumber\\
\tq_i &:& (0,0,\ha)^0 \nonumber\\
(\p_{q})_\a &:& (\ha,0,0)^{+1} \\
(\pb_{q} )_\da &:& (0,\ha,0)^{-1}
\nonumber\\
(\p_{\tq} )_\a &:& (\ha,0,0)^{+1} \nonumber\\ 
(\pb_{\tq})_\da &:& (0,\ha,0)^{-1}\ .
\nonumber
\end{eqnarray}
The complete N=2 SYM action is: 
\begin{equation}
{S}={S}_{YM}+{S}_{matter} \ ,\label{azionetotale}
\end{equation}
where ${S}_{YM}$ is given by (\ref{SYMN2pg}), and
\begin{eqnarray}\label{SYMN2mat}
{S}_{matter}&=&\frac{1}{g^2}\tr_{m}\xint \Big( \ha \tq^i D_\m
D^\m q_i + 2\tq^i\lb_{i\da}(\pb_q)^\da - 2q^i
\lb_{i\da}(\pb_{\tq})^\da 
\nonumber\\ 
&-& \ha\tq^i\l_{i\a}(\p_q)^\a - \ha q^i\l_{i\a}(\p_{\tq})^\a +
(\p_{\tq})^\a(\s^\m)_{\a\da}D_\m (\pb_q)^\da 
\nonumber\\ 
&-& (\pb_{\tq})_\da(\sb^\m)^{\da\a}D_\m (\p_q)_\a +
\frac{1}{8}(\p_{\tq})^\a \fb(\p_q)_\a - 2(\pb_{\tq})^\da\f(\pb_q)_\da
\nonumber\\ 
&+&\frac{1}{16}\tq^i \left\{ \f,\fb \right\} q_i -
\frac{1}{32}\tq^i q_i \tq^j q_j \Big)\ .
\label{smatter}\end{eqnarray}
In the previous expression, $\tr_{m}$ is the Trace over the 
matter representation of the gauge group.

The (nonlinear) supersymmetry transformations of the matter fields are: 
\begin{eqnarray}\label{susytransfN2mat}
\delta q_i&=&\rd\ve_{ji}\xi^{\a j}(\p_q)_\a + \rd\ve_{ji}\xib^{\da
j}(\pb_q)_\da\nonumber\\
\delta {\tq}_i&=&\rd\ve_{ji}\xi^{\a j}(\p_{\tq})_\a +
\rd\ve_{ji}\xib^{\da j}(\pb_{\tq})_\da\nonumber\\
\delta(\p_q)_\g&=&\rd\ve_{\g\a}\xi^{\a j}\f q_j
+\frac{1}{\rd}\xib^{\da j}(\s^\n)_{\g\da}{\rm D}_\n q_j \nonumber\\
\delta(\pb_q)_\dg&=&-\frac{1}{\rd}\xi^{\a j}(\s^\n)_{\a\dg}{\rm
D}_\n q_j - \frac{\rd}{16}\ve_{\dg\da}\xib^{\da j}\fb q_j \\
\delta(\p_{\tq})_\g&=&-\rd\ve_{\g\a}\xi^{\a j}\f \tq_j
+\frac{1}{\rd}\xib^{\da j}(\s^\n)_{\g\da}{\rm D}_\n \tq_j \nonumber\\
\delta(\pb_{\tq})_\dg&=&-\frac{1}{\rd}\xi^{\a j}(\s^\n)_{\a\dg}{\rm
D}_\n \tq_j + \frac{\rd}{16}\ve_{\dg\da}\xib^{\da j}\fb \tq_j\ ,
\nonumber
\end{eqnarray}
and the matter action (\ref{SYMN2mat}) is susy invariant: 
\bea
\delta {S}_{matter}=0\ ,
\eea
so that, finally, one has
\bea
\delta {S}=\delta ({S}_{YM}+{S}_{matter})=0\ .
\eea

\section{Introducing the twist: the pure N=2 SYM theory}

As we said, the global symmetry group for N=2 SYM in four dimensions 
is given by $H$ \equ{H}, and the total number of generators, 
including supersymmetry, are:
\begin{center}
\begin{tabular}{|l|c|c|c|c|}
\hline
&&&&\\
& $SU(2)_{L}\times SU(2)_{R}$ & Susy & $SU(2)_I$ & $U(1)$ \\ 
&&&&\\
\hline
&&&&\\
generators&$P_\m(4)\,,\,M_{\m\n}(6)$&$Q_{i\a}(4)\,,\,
\ov Q_{i\da}(4)$&$T^i_j(3)$&${\cal R}(1)$ \\ 
&&&&\\ 
\hline
\end{tabular}
\end{center} 
The nonvanishing algebraic relations are
\begin{eqnarray}\label{algtot}
\left[M_{\mu\nu},M_{\rho\sigma}\right]&=&-i(\eta_{\mu\rho}M_{\nu\sigma}-
\eta_{\mu\sigma}M_{\nu\rho}-\eta_{\nu\rho}M_{\mu\sigma}+
\eta_{\mu\sigma}M_{\mu\rho}) 
\nonumber\\ 
\left[M_{\mu\nu},P_\rho\right]&=&i(\eta_{\nu\rho}P_\mu-
\eta_{\mu\rho}P_\nu) 
\nonumber\\ 
\left[M_{\m\n},Q^i_\a \right]&=&- (\s_{\m\n})_\a^{\ \b} Q^i_\b 
\nonumber\\ 
\{Q^i_\a, \ov Q^j_\db \}&=&2 \s^\m_{\a\db} P_\m \delta^{ij}
\\
\left[T_i^j,Q_{k\a}\right]&=&-\ha(\delta_k^jQ_{i\a}-\ha\delta_i^jQ_{k\a})
\nonumber\\
\left[T_i^j,T_k^l\right]&=&\ha(\delta_i^l T_k^j-\delta_k^j T_i^l)
\nonumber\\
\left[{\cal R}, Q_{i\a}\right]&=&Q_{i\a}\ ,
\nonumber
\end{eqnarray}
and their hermitian conjugates. Remember that the $SU(2)_{I}$ 
generators are traceless: $T_i^i=0$, hence only three of them are 
independent.

It is now convenient to rearrange the Lorentz and translations 
generators $M_{\m\n}$ and $P_\m$ as follows: 
\bea
J_{\a\b}:=\ha(\s^{\m\n})_{\a\b}M_{\m\n}
\ \ ;\ \ 
\ov{J}_{\adot\bdot}:=\ha(\ov{\s}^{\m\n})_{\adot\bdot}M_{\m\n}
\ \ ;\ \ 
P_{\a\db}:=(\s^\m)_{\a\db}P_\m\ ,
\eea
exploiting the isomorphism $\r$ between the 
Minkowski space $M_{4}$ and the space $H(2,C)$ of $2\times 2$ hermitian 
matrices:
\begin{eqnarray}
    \r:\ M_4\rightarrow H&\ ,\ & \r(x_\m)=x_\m \s^\m\\
\r^{-1}:\ H\rightarrow M_4&\ ,\ & \r^{-1}(h)=\ha 
\Tr\left[h\sb^\m\right]\ .
\end{eqnarray}

The twisting procedure, introduced by Witten in 
\cite{Witten:1988ze,Witten:1994ev}, simply 
consists into a redefinition of the internal group indices $i$ as 
lefthanded spinorial indices~$\a$:
\bea
i \stackrel{twist}{\longrightarrow} \a\ .
\eea
This is possible thanks to the fact that both the spinorial indices 
$\{\a,\da\}$ and the susy index $i$ run from 1 to 2. The Lorentz 
group generators $J_{\a\b}$ are correspondingly redefined through a linear 
combination $J'_{\a\b}$ with the $SU(2)_{I}$ internal group generators, which, 
after the twist, are written as $T_\a^\b$:
\begin{equation}\label{twist}
J'_{\a\b}:=J_{\a\b} + kT_{\a\b}\ ,\label{jprimo}
\end{equation}
where $k$ is a constant to be fixed by requiring that 
$\left[J',J'\right]=\left[J,J\right]$. Since both $J$ and $T$ are 
symmetric in $(\a,\b)$, the same holds also for $J'$. Notice that 
lefthandedness is a possibility, the twist defined through the 
identification of $i$ and $\adot$ being equally legitimate.

If $SU(2)_{L}$ is the group associated to the generators $J_{\a\b}$, 
the redefinition \equ{jprimo} corresponds to twisting the Lorentz group 
$SU(2)_L\times SU(2)_R$ into $SU(2)'_L\times SU(2)_R$, where 
$SU(2)'_L$ is the diagonal sum of $SU(2)_L$ and $SU(2)_I$.

The new, twisted, global symmetry group $H'$ is
\bea
H \stackrel{twist}{\longrightarrow} H'=SU(2)'_L\times SU(2)_R\times 
U(1)\ .
\label{hprime}\eea
The supersymmetry charges become:  
\bea
Q_{i\a}\stackrel{twist}{\longrightarrow} Q_{\b\a} \quad 
\mbox{and}
\quad \ov Q_{i\da}\stackrel{twist}{\longrightarrow}\ov Q_{\b\da}\ .
\eea
The twisted supercharges under $H'$ transform as 
$Q_{\b\a}=(0,0)^{+1}\oplus(1,0)^{+1}$ and $\ov 
Q_{\b\da}=(\ha,\ha)^{-1}$\ , or, more explicitly, the four 
supercharges $Q_{\b\a}$ under 
the twist can be rearranged into a scalar $\delta_{W}$ and an 
anti-selfdual tensor $\delta_{\m\n}$, while the other four $\ov Q_{\b\da}$ 
become a vector operator $\delta_{\m}$:
\begin{eqnarray}\label{twistcariche}
Q_{\b\a}&\stackrel{twist}{\longrightarrow}& \dw
:=\frac{1}{\rd}\ve^{\a\b}Q_{\b\a}\,\oplus\,\dmn:=
\frac{1}{\rd}(\s_{\m\n})^{\a\b}Q_{\b\a}\nonumber\\
\ov Q_{\b\da}&\stackrel{twist}{\longrightarrow}& \dmu
:=\frac{1}{\rd} \ov Q_{\b\da}(\ov\s_\m)^\da_\b\ ,
\end{eqnarray}
and $\delta_{\m\n}$ is selfdual
\bea
\dmn=\tilde \dmn=\ha \ve_{\m\n\r\s}\delta^{\r\s}\ .
\eea
The subalgebra formed by the eight twisted supercharges 
$\delta_{W},\delta_{\m},\delta_{\m\n}$ and the  ${\cal R}$ symmetry, 
reads:
\begin{eqnarray}\label{algtwistata}
\{\dw,\dw\}&=&2\dw^2=0
\nonumber\\
\{\dw,\dmu\}&=&\d_\m
\nonumber\\
\{\dmu,\dnu\}&=&0\\
\{\dmu,\drs\}&=&-\left(\ve_{\m\r\s\n}\d^\n+g_{\m\r}\d_\s-
g_{\m\s}\d_\r\right)
\nonumber\\
\{\dw,\dmn\}&=&0
\nonumber\\
\left[{\cal R}, \dw\right]&=&+\dw
\nonumber\\
\left[{\cal R}, \dmu\right]&=&-\dmu
\nonumber\\
\left[{\cal R}, \dmn\right]&=&+\dmn\ ,
\nonumber
\end{eqnarray}
where $g_{\m\n}=diag(+,+,+,+)$ is the euclidean flat space metric.

A few remarks are in order: 
\begin{enumerate}
\item 
The operator $\dw$, which coincides with the ``fermionic symmetry'' 
introduced by Witten in \cite{Witten:1988ze,Witten:1994ev}, is nilpotent
\bea
\dw^{2}=0\ .
\eea
In the 
Wess Zumino gauge its realization is nonlinear, as we shall see, and 
$\dw$ will turn out to be nilpotent $modulo$ (field-dependent) 
gauge transformations and field equations, as usual in 
supersymmetry algebras.
\item
The operators $\dw$ and $\dmu$ form a subalgebra which closes on 
translations. This is a common, and somehow defining, feature of 
topological models \cite{Birmingham:1991ty}, 
and remarkably suggests that the twist has 
deeply to do with topological quantum field theories and their
algebraic structure. In fact the common feature of all topological 
field theories, is the existence of three operators $\delta,\dmu,\d_\m$ 
satisfying the following algebra \cite{Piguet:1995er}
\bea
\delta^2=0\quad ,\quad \{\delta,\dmu\}=\d_\m \quad ,\quad \{\dmu,\dnu\}=0\ .
\eea
In other words, it is not surprising at all that, twisting N=2 SYM, a topological
quantum field theory is recovered.
\item
The twist does not change the mass dimensions of the supersymmetry 
charges, which is $\ha$. The ${\cal R}$-charge is $+1$ for $\dw$ and $\dmn$, 
and $-1$ for the vector symmetry $\dmu$.

\item The following table summarizes the effect of the twist on the 
global group, its generators and on the supersymmetry charges: 
\begin{center}
\begin{tabular}{|c|c|c|c|c|c|c|c|}
\hline
&\multicolumn{3}{|l|}{ \ \ \ \ \ \ \ UNTWISTED}&&\multicolumn{3}{|l|}{ \ \ \ \ \ \ \ TWISTED}\\
\cline{1-4}\cline{7-8}
\hline
group& $SU(2)_L$ & $SU(2)_R$ & $SU(2)_I$&  && $SU(2)'_L$ & $SU(2)_R$\\ 
\cline{1-4}\cline{7-8}
generators & $J_{\a\b}(3)$ & $\ov J_{\da\db}(3)$ & $T_{ij}(3)$& 
&& $J'_{\a\b}(3)$ & $\ov J_{\da\db}(3)$\\
\cline{1-4}\cline{6-8}
$Q_{i\a}$ & 1/2 & $0$ & 1/2&  & $\dw$ & $0$ & $0$ \\ 
\cline{1-4}\cline{6-8}
\multicolumn{4}{|l|}{ }&  & $\dmn$ & $1$ & $0$\\
\cline{1-4}\cline{6-8}
$\ov Q_{i\da}$ & $0$ & 1/2 & 1/2 && $\dmu$ & 1/2 & 1/2 \\
\hline
\end{tabular}
\end{center}
\end{enumerate}

\subsection{Twisted fields}

The fields belonging to the YM multiplet concerned by the twist 
are the fermionic fields $\l^{i\a}(x)$ and $\ov\l_{i\adot}(x)$, $i.e.$ those 
carrying the internal supersymmetry index $i$, which, like the 
supercharges $Q_{i\a}$ and $\ov Q_{i\adot}$, are twisted as follows
\begin{eqnarray}
\l^{i\a}\rightarrow\l_{\b\a}(\ha,0,\ha)^{-1}&\rightarrow&\eta(0,0)^{-1}\oplus\chi_{\m\n}(1,0)^{-1}
\\
\ov\l_{i\adot}\rightarrow\lb_{\b\da}(0,\ha,\ha)^{+1}&\rightarrow&\p_\m(\ha,\ha)^{+1}\ .
\end{eqnarray}
The field $\l(x)$ is twisted into a scalar field $\eta(x)$ and an antiselfdual 
antisymmetric tensor $\chi_{\mu\nu}(x)$, while $\ov\l(x)$ yields a vector 
field $\psi_{\mu}(x)$: 
\begin{eqnarray}\label{twistcampi}
\l_{\b\a}&\rightarrow& \eta:=\ve^{\a\b}\l_{[\b\a]}\,\oplus\,\chi_{\m\n}:=
\frac{1}{4}(\s_{\m\n})^{\a\b}\l_{(\b\a)}\nonumber\\
\lb_{\b\da}&\rightarrow& \p_\m:=\lb_{\b\da}(\ov\s_\m)^\da_\b\ ,
\end{eqnarray}
with
\bea
\chi_{\m\n}=\tilde \chi_{\m\n}=\ha \ve_{\m\n\r\s}\chi^{\r\s}\ .
\eea
The bi-spinor $\l_{\b\a}(x)$ is thus decomposed into its symmetric and 
antisymmetric part: $\l_{\b\a}=\ha(\l_{[\b\a]}+\l_{(\b\a)})$.

Summarizing, the effect of the twist on the fields of the YM multiplet is
\bea
(A_\m,\l_{i\a},\lb_{i\da},\f,\fb)
\stackrel{twist}{\longrightarrow}
(A_\m,\p_\m,\chi_{\m\n},\eta,\f,\fb)\ ,
\eea
which, not by chance, coincides with the field content of the 
Donaldson-Witten topological QFT
\cite{Witten:1988ze,Donaldson:1983wm}.

\subsection{Twisted action}

The twisting procedure changes the action \equ{sym} accordingly. It is 
important to stress that the twist, as far as quantum 
fields are concerned, is simply a linear 
rearrangement, which does not modify the path 
integral defining the functional generators. The partition function is 
not affected by the twist, hence the two theories, the untwisted and 
the twisted one, are completely equivalent. This means, in 
particular, that the physical observables should be the same in the 
two theories, that the finiteness properties must be preserved, and that 
the susy invariance should reflect into invariance under the twisted 
operators $\dw$, $\delta_{\m}$ and $\delta_{\mu\nu}$. 

In order to 
verify this latter property, let us twist the pure N=2 SYM action~\equ{sym}:
\begin{eqnarray}\label{TYM}
S_{YM}
\stackrel{twist}{\longrightarrow}
S_{TYM}&=&
\frac{1}{g^2}{\rm Tr}\xint \Big(\ha F_{\m\n}^{+}F^{+\m\n}-
\chi^{\m\n}\left(D_\m \p_\n -D_\n\p_\m\right)^+\nonumber\\
&+&\eta D_\m \p^\m-\ha\fb D_\m D^\m \f +\ha\ov\f \{ \p^\m,\p_\m \} \\
&-&\ha\f\{ \chi^{\m\n},\chi_{\m\n}\}-\frac{1}{8}\left[\f,\eta \right]\eta-
\frac{1}{32}
\left[ \f ,\fb \right] \left[ \f ,\fb \right] \Big)\ . \nonumber
\end{eqnarray}
It is evident that the twisted N=2 SYM theory coincides with the Topological Yang-Mills 
(TYM) theory, as expected \cite{Witten:1988ze,Witten:1994ev}.

Of course, under the gauge transformations
\begin{eqnarray}\label{g-transf}
\delta _\epsilon ^gA_\mu &=&-D_\mu \epsilon  \\
\delta _\epsilon ^g\lambda &=&\left[ \epsilon ,\lambda \right]
\;,\;\;\;\;\lambda =\chi ,\psi ,\eta ,\phi ,\overline{\phi } \ , \nonumber
\end{eqnarray}
$\epsilon(x)$ being the local infinitesimal gauge parameter, the action $S_{TYM}$ is invariant
\begin{equation}\label{g-inv}
\delta _\epsilon ^g S_{TYM}=0.  
\end{equation}
As we already said, we expect that $S_{TYM}$ keeps memory 
of susy invariance through its invariance under the twisted operators 
$\dw$, $\delta_{\m}$ and $\delta_{\mu\nu}$. 
Notice that it would have been a very difficult task to identify 
{\it a priori} the symmetries of the TYM action \equ{TYM}. 
This, on the contrary, turns 
out to be quite natural thanks to the twisting procedure.
Before verifying that the twisted operators are indeed symmetries of 
the TYM action, we have to write down their action on the twisted 
fields.

\subsection{Twisted supersymmetry transformations on twisted fields}

Recalling the definition of the twisted operators 
(\ref{twistcariche}) and of the twisted fields (\ref{twistcampi}), 
after a little algebra, one gets
\begin{description}
    \item{\bf $\dw$ transformations on twisted fields}
\begin{eqnarray}\label{dw}
\dw A_\m &=& \p_\m \nonumber \\
\dw\p_\m &=& -D_\m \f \nonumber \\
\dw\phi &=& 0\nonumber \\
\dw\chi_{\m\n} &=& F_{\m\n}^{+} \\
\dw\ov\f &=& 2\eta \nonumber \\
\dw\eta &=& \frac{1}{2}\left[\f,\ov\f\right] \nonumber
\end{eqnarray}
\item{\bf $\dmu$ transformations on twisted fields}
\begin{eqnarray}\label{dmu}
\dmu A_\n &=&\ha\chi_{\m\n}+\frac{1}{8}g_{\m\n}\eta \nonumber\\
\dmu\p_\n &=&F_{\m\n}-\ha F_{\m\n}^{+}-\frac{1}{16}g_{\m\n}\left[\f,\ov\f\right]
\nonumber\\
\dmu\eta &=&\ha D_\m\ov\f \\
\dmu\chi_{\s\tau} &=&
\frac{1}{8}\left(\ve_{\m\s\tau\n}D^\n \ov\f+
g_{\m\s}D_\tau\ov\f-g_{\m\tau}D_\s\ov\f\right)\nonumber\\
\dmu\f &=&-\p_\m \nonumber \\
\dmu\ov\f &=&0\nonumber
\end{eqnarray}
\item{\bf $\dmn$ transformations on twisted fields}
\begin{eqnarray}\label{dmn}
\dmn A_\s &=&-\left(\ve_{\m\n\s\tau}\p^\tau+g_{\m\s}\p_\n-g_{\n\s}\p_\m \right)
\nonumber \\
\dmn \p_\s &=&-\left(\ve_{\m\n\s\tau}D^\tau\f +
g_{\m\s}D_\n\f-g_{\n\s}D_\m\f\right)
\nonumber \\
\dmn \f &=&0 \nonumber \\
\dmn \ov\f &=&8\chi_{\m\n} \\
\dmn \eta &=&-4F_{\m\n}^+  \nonumber \\
\dmn \chi_{\s\tau} &=&\frac{1}{8}\left(\ve_{\m\n\s\tau}+
g_{\m\s}g_{\n\tau}-g_{\m\tau}g_{\n\s}\right)\left[\f,\ov\f\right] 
\nonumber \\
&&+\left(F_{\m\s}^+ g_{\n\tau}-F_{\n\s}^+ g_{\m\tau}-F_{\m\tau}^+ g_{\n\s}+
F_{\n\tau}^+g_{\m\s}\right) \nonumber \\
&&+\left(\ve _{\m\n\s}^{\ \ \ \ \a}F_{\tau\a}^+ -\ve_{\m\n\tau}^{\ \ \ \ \a }
F_{\s\a}^+ +
\ve_{\s\tau\m}^{\ \ \ \ \a}F_{\n\a}^+ -\ve_{\s\tau\n}^{\ \ \ \ \a }F_{\m\a}^+
\right) \ .
\nonumber
\end{eqnarray}
The following tables summarize the quantum numbers of the twisted 
fields 
\begin{center}
\begin{tabular}{|c|c|c|c|c|c|c|}
\hline
twisted fields& $A_\mu $ & $\chi _{\mu \nu }$ & $\psi _\mu $ & 
$\eta $ & $\phi $ & $\fb$ \\ \hline
$\mathrm{dim}.$ & $1$ & $3/2$ & $3/2$ & $3/2$ & $1$ & $1$\\ 
\hline
$\mathcal{R}-\mathrm{charge}.$ & $0$ & $-1$ & $1$ & $-1$ & $2$ & $-2$ \\ 
\hline
$\mathrm{statistics}$ & $comm$ & $ant$ & $ant$ & $ant$ & $comm$ & $comm$ \\ 
\hline
\end{tabular}
\end{center}
and of the twisted operators
\begin{center}
\begin{tabular}{|c|c|c|c|}
\hline
twisted operators&$\dw$ & $\dmu$ & $\dmn$ \\ \hline
$\mathrm{dim}.$ & $1/2$ & $1/2$ & $1/2$ \\ 
\hline
$\mathcal{R}-\mathrm{charge}.$ & $1$ & $1$ & $-1$\\ 
\hline
$\mathrm{statistics}$ & $ant$  & $ant$  & $ant$\\ \hline
\end{tabular}
\end{center}
\end{description}
where $comm$ and $ant$ stand for $commuting$ and $anticommuting$ 
respectively.

Long but straightforward calculations confirm that, indeed, 
the twisted operators are symmetries of the twisted action: 
\bea
\dw S_{TYM}=
\dmu S_{TYM}=
\delta_{\mu\nu} S_{TYM}=0\ .
\eea
It is important to stress that the fermionic, nilpotent, Witten's 
$\dw$ symmetry does not completely fix the coefficients of every 
term appearing  in $S_{TYM}$. In other words, 
$S_{TYM}$ {\it is not} the most general action invariant under 
$\dw$. In order to fix completely all the terms by means of {\it a 
unique} coupling constant, the role of the vector $\dmu$ symmetry is 
crucial. On the other hand, the three $\dmn$ twisted symmetries 
are automatically satisfied, therefore, under this respect, they seem 
to be redundant.

\subsection{Twisted algebra}

Let us see what becomes the twisted supersymmetry algebra in the 
Wess-Zumino gauge, where the symmetries are nonlinearly realized.
The following algebraic relations hold: 

\begin{equation}
\dw^2=\delta_\f^g\;+\;(\textrm{field equations}) \label{dw2}\ ,
\end{equation}
where $\delta_\f^g$ is a gauge transformation whose gauge parameter 
is the field $\f(x)$. The operator $\dw$ is therefore {\it on shell} nilpotent in the 
space of gauge invariant local functionals. 
The cohomology in this constrained functional space defines 
the so called Witten observables \cite{Witten:1988ze,Witten:1994ev};

\begin{equation}\label{dmudnu}
\left\{ \dmu ,\dnu \right\}=-\frac{1}{8}g_{\m\n}\delta_\fb^g\;+
(\textrm{field equations}),
\end{equation}
where $\delta_\fb^g$ is a field dependent gauge transformation, with 
the field $\fb(x)$ as gauge parameter;

\begin{equation}\label{dwdmu}
\left\{ \dw,\dmu \right\}=\d_\m +\delta_{A_\m}^g+(\textrm{field 
equations}),
\end{equation}
where $\delta_{A_\m}^g$ is a field dependent gauge transformation, with 
the field $A_\m(x)$ as gauge parameter.

Finally, the algebraic relations involving $\delta_{\mu\nu}$ are:

\begin{equation}\label{dwdmn}
\left\{\dw,\dmn\right\}=(\textrm{gauge 
transformation})+(\textrm{field equations}) ;
\end{equation}\\

\begin{equation}\label{dmndrs}
\left\{\dmn,\drs\right\}=(\textrm{gauge 
transformation})+(\textrm{field equations}) ;
\end{equation}\\

\begin{eqnarray}\label{dmudrs}
\left\{ \dmu ,\drs\right\}&=&
-(\ve_{\m\r\s\n}\d ^\n +g_{\m\r}\d_\s -g_{\m\s}\d_\r)
\\
&&
+(\textrm{gauge 
transformation})+(\textrm{field equations}) .\nonumber
\end{eqnarray}
The above algebraic structure is typical of the supersymmetry in the 
Wess-Zumino gauge. Two kind of obstructions to the closure of the 
algebra on translations occur: field equations and {\it field dependent} gauge 
transformations. The canonical way to proceed 
(see, for instance, \cite{West:1990tg}), is to take care of 
the first type of obstructions, namely the field equations, 
introducing auxiliary fields, whose transformations coincide with 
the field equations. Still, the other kind of obstruction, namely the field 
dependent gauge transformations, remains, and the algebra is
{\it open}, needing an infinite number of external fields. This 
problem has been exhaustively treated in \cite{Breitenlohner:1985kh}, where 
the non-renormalizability of theories with extended supersymmetry is 
discussed. The problem has nonetheless been solved, turning the 
situation the other way around \cite{Maggiore:1994dw,Maggiore:1994xw}, 
as we shall see.

\section{The twisted theory: N=2 SYM coupled to matter}

Let us now apply the twisting procedure, described in the previous 
section, to the complete N=2 SYM theory, coupled to matter. Besides 
the pure YM multiplet, belonging to the adjoint representation of the 
gauge group, the field content of the theory is completed by the 
hypermultiplet, in a generic 
representation of the gauge group. The global symmetry group does not 
change, and the twist goes the same way: 
\bea
H \stackrel{twist}{\longrightarrow} H'\ ,
\eea
where $H$ and $H'$ are defined in \equ{H} and \equ{hprime} respectively.

In this section, we shall find out the twisted matter 
fields, the complete twisted action, the 
twisted operators and the corresponding twisted algebra. We shall 
moreover verify that the twisted operators are still symmetries of the 
twisted theory. The result should not be taken for granted, since the 
topological character of the twist is spoiled by the introduction of 
matter, and therefore we do not expect that the twisted theory
is topological. Hence, the algebraic topological structure, which we 
shall find for a non-topological field theory, comes somehow as a surprise. 

\subsection{Twisted hypermultiplet}

The matter hypermultiplet is 
$(q_{i},\tq_{i},\psi_{q},\pb_{q},\p_{\tq},\pb_{\tq})$. Only the 
bosonic fields $q_{i}(x)$ and $\tq_{i}(x)$, which have a nonvanishing $SU(2)_I$ 
quantum number, will be twisted, the other fields remaining unchanged.
The action of the twist is as follows, and we rename the fields in 
order to simplify notations: 
\begin{eqnarray}\label{twistiper}
q_{i}\stackrel{twist}{\longrightarrow}
q_\a(0,0,\ha)^0 &\rightarrow&\H_\a(\ha,0)^0\nonumber\\
\tq_{i}\stackrel{twist}{\longrightarrow}
\tq_\a(0,0,\ha)^0 &\rightarrow&\Hb_\a(\ha,0)^0\nonumber\\
(\p_q)_\a (\ha,0,0)^{+1}&\rightarrow& u_\a(\ha,0)^{+1}\\
(\pb_q)_\da (0,\ha,0)^{-1}&\rightarrow& v_\da(0,\ha)^{-1}\nonumber\\
(\p_{\tq})_\a (\ha,0,0)^{+1}&\rightarrow& \ub_\a(\ha,0)^{+1}\nonumber\\
(\pb_{\tq})_\da (0,\ha,0)^{-1}&\rightarrow&\vb_\da(0,\ha)^{-1}\ .
\nonumber
\end{eqnarray}
Notice that, while in the pure  N=2 SYM the twist gets rid of 
the spinorial fields, this does not happen for the hypermultiplet, 
whose twisted version, on the contrary, is entirely formed by spinors.

\subsection{Twisted matter action}

Twisting the matter N=2 SYM action \equ{smatter}, we get
\bea
S_{matter}
\stackrel{twist}{\longrightarrow}
S_{Tmatter}\ ,
\eea
with
\begin{eqnarray}\label{matterat}
S_{Tmatter}&=&
\frac{1}{g^2}{\rm Tr_m}\xint \Big(\ha \Hb^\g D_\m D^\m \H_\g+
\Hb^\g(\s^\m)_{\g\dg}\p_\m v^\dg 
\\
&&
-\vb_\dg(\sb^\m)^{\dg\g}\p_\m \H_\g
+\frac{1}{8}\Hb^\g\eta u_\g+
\frac{1}{8}\Hb^\g (\s^{\m\n})_{\g\b}\chi_{\m\n} u^\b+
\frac{1}{8}\ub^\g\eta \H_\g \nonumber\\
&&-\frac{1}{8}\ub^\g (\s^{\m\n})_{\g\b}\chi_{\m\n} \H^\b+
\ub^\g (\s^\m)_{\g\dg}D_\m v^\dg-\vb_\dg (\sb^\m)^{\dg\g}D_\m u_\g
\nonumber \\
&&+
\frac{1}{8}\ub^\g \fb u_\g
-2 \vb^\dg \f v_\dg + \frac{1}{16}\Hb^\g\{\f,\fb\}\H_\g-
\frac{1}{32}\Hb^\g \H_\g \Hb^\delta \H_\delta\Big)\nonumber
\end{eqnarray}

\subsection{Twisted supersymmetry transformations on twisted fields}

The action of the twisted operators $(\dw,\dmu,\dmn)$ on  the twisted 
matter fields, is as follows
\begin{description}
    \item{\bf $\dw$ transformations on twisted matter fields}
\begin{eqnarray}\label{iperdw}
\dw \H_\g&=&\frac{1}{\rd}\ve^{\a\b}Q_{\b\a}\H_\g=
\frac{1}{\rd}\ve^{\a\b}(\rd u_\a \ve_{\b\g})=u_\g\nonumber\\
\dw \Hb_\g&=&\ub_\g\nonumber\\
\dw u_\g&=&\frac{1}{\rd}\ve^{\a\b}(\rd \ve_{\g\a}\f\H_\b)=+\f\H_\g\nonumber\\
\dw \ub_\g&=&-\f\Hb_\g\\
\dw v_\dg&=&\frac{1}{\rd}\ve^{\a\b}
\left(-\frac{1}{\rd}(\s^\n)_{\a\dg}D_\n\H_\b\right)=
-\ha(\s^\n)_{\a\dg}D_\n\H^\a\nonumber\\
\dw \vb_\dg&=&-\ha(\s^\n)_{\a\dg}D_\n\Hb^\a\nonumber
\end{eqnarray}
\item{\bf $\dmu$ transformations on twisted matter fields}
\begin{eqnarray}\label{iperdmu}
\dmu \H_\g&=&\frac{1}{\rd}(\sb_\m)^{\da\b}\left(\rd v_\da\ve_{\b\a}\right)=
(\s_\m)_{\g\da}v^\da\nonumber\\
\dmu \Hb_\g&=&(\s_\m)_{\g\da}\vb^\da\nonumber\\
\dmu u_\g&=&
\frac{1}{\rd}(\sb_\m)^{\da\b}\left(\frac{1}{\rd}(\s_\n)_{\g\da}D^\n\H_\b\right)=
\ha D_\m\H_\g-\ha(\s_{\m\n})_\g^{\ \b}D^\n\H_\b \nonumber\\
\dmu \ub_\g&=&\ha D_\m\Hb_\g-\ha(\s_{\m\n})_\g^{\ \b}D^\n\Hb_\b \\
\dmu v_\dg&=&\frac{1}{\rd}(\sb_\m)^{\da\b}
\left(\frac{\rd}{16}\ve_{\dg\da}\fb\H_\b\right)
=-\frac{1}{16}(\sb_\m)_{\dg\b}\fb\H^\b\nonumber\\
\dmu \vb_\dg&=&\frac{1}{16}(\sb_\m)_{\dg\b}\fb\Hb^\b\nonumber
\end{eqnarray}
\item{\bf $\dmn$ transformations on twisted matter fields}
\begin{eqnarray}\label{iperdmn}
\dmn \H_\g&=&\frac{1}{\rd}(\s_{\m\n})^{\a\b}\left(\rd u_\a \ve_{\b\g}\right)=
-(\s_{\m\n})_\g^{\ \a}u_\a\nonumber\\
\dmn \Hb_\g&=&-(\s_{\m\n})_\g^{\ \a}\ub_\a\nonumber\\
\dmn u_\g&=&\frac{1}{\rd}(\s_{\m\n})^{\a\b}\left(\rd \ve_{\g\a}\f\H_\b\right)=
(\s_{\m\n})_\g^{\ \b}\f\H_\b\nonumber\\
\dmn \ub_\g&=&-(\s_{\m\n})_\g^{\ \b}\f\Hb_\b\\
\dmn v_\dg&=&\frac{1}{\rd}(\s_{\m\n})^{\a\b}
\left(-\frac{1}{\rd}(\s^\l)_{\a\dg}D_\l \H_\b \right)=
\ha(\s_{\m\n})_\b^{\ \a}(\s_\l)_{\a\dg}D^\l \H^\b \nonumber\\
&=&\ha(\s_\m)_{\b\dg}D_\n\H^\b-\ha(\s_\n)_{\b\dg}D_\m\H^\b-
\ha\ve_{\m\n\l\tau}(\s^\tau)_{\b\dg}D^\l\H^\b\nonumber\\
&=&\ha\left[(\s_\m)_{\b\dg}D_\n\H^\b-(\s_\n)_{\b\dg}D_\m\H^\b\right]^+
\nonumber\\
\dmn \vb_\dg&=&\ha\left[(\s_\m)_{\b\dg}D_\n\Hb^\b-
(\s_\n)_{\b\dg}D_\m\Hb^\b\right]^+ \ .
\nonumber
\end{eqnarray}
\end{description}
The following table summarizes the quantum number and the statistics of the 
twisted matter fields:
\begin{center}
\begin{tabular}{|c|c|c|c|c|c|c|}
\hline
twisted hypermultiplet& $\H$ & $\Hb$ & $u$ & $\ub$ & $v$ & $\vb$ \\ \hline
$\mathrm{dim}.$ & $1$ & $1$ & $3/2$ & $3/2$ & $3/2$ & $3/2$\\ 
\hline
$\mathcal{R}-\mathrm{charge}.$ & $0$ & $0$ & $+1$ & $+1$ & $-1$ & $-1$ \\ 
\hline
$\mathrm{statistics}$ & $comm$. & $comm$. & $ant.$ & $ant.$ & $ant.$ & $ant.$ \\ 
\hline
\end{tabular}
\end{center}

\subsection{Twisted algebra}

Once we have the twisted action 
(\ref{matterat})
and the twisted field transformations \equ{iperdw} and \equ{iperdmu}, 
we can verify that the algebra formed by the 
twisted operators $(\dw,\dmu,\dmn)$ is the same of the pure gauge 
case, {\it i.e.} it is a 
topological, supersymmetric algebra which closes on translations, {\it 
modulo} gauge dependent field transformations and equations of 
motion. Notice that the fields on which the gauge transformations 
depend, are the same as in the pure gauge case.

The complete N=2 SYM theory contains interactions terms between the two 
supermultiplets, the pure gauge and the matter one. Hence, the field 
equations of motion for the vector supermultiplet change. In order to 
preserve the algebra, which depends on the field equations, 
we must modify the 
transformations of the gauge multiplet. Let us see how 
this can be done. The field equations appearing in the algebra as 
obstructions, are those concerning the fields $\eta(x), \chi(x)$ and 
$\psi(x)$, 
which appear in the interaction terms of the complete action. 
Their transformations under the twisted operators are those to be 
changed. Let us see in detail how, for example, the 
transformation $\dw\chi_{\mu\nu}$ must be modified.

Since
\begin{eqnarray}\label{proc.x.chi.tr}
\dw^2\chi_{\m\n} &=&\left[\f,\chi_{\m\n}\right] -
g^2\frac{\delta S_{TYM}}{\delta \chi ^{\m\n}} \nonumber\\
&=&\left[\f,\chi_{\m\n}\right] -\left[\f,\chi_{\m\n}\right]+
\left(D_\m \p_\n-D_\n\p_\m\right)^+ \\
&&-\frac{1}{8}\Hb^\g(\s^{\m\n})_{\g\b}u^\b-
\frac{1}{8}\ub^\g(\s^{\m\n})_{\g\b}\H^\b\ , \nonumber
\end{eqnarray}
it must be
\begin{equation}\label{newchitransf}
\dw\chi_{\m\n}=F^+_{\m\n}-
\frac{1}{8}\Hb^\g(\s_{\m\n})_{\g\b}\H_\g\ .
\end{equation}
Analogously, by analyzing the whole set of algebraic relations, 
from \equ{dw2} to (\ref{dmudrs}), we can infer the modified
transformations of the fields belonging to the YM multiplet, when coupled to the 
matter hypermultiplet: 
\begin{eqnarray}\label{lastnewtransf}
\dw\chi_{\m\n}&=&F^+_{\m\n}-
\frac{1}{8}\Hb^\g(\s_{\m\n})_{\g\b}\H_\g\nonumber\\
\dmu \p_\n&=&F^-_{\m\n}-\frac{1}{16}g_{\m\n}\left[\f,\fb\right]+
\frac{1}{16}\Hb^\a(\s_{\m\n})_{\a\b}\H^\b\nonumber\\
\dmn \eta&=&-4F^+_{\m\n}+\ha\Hb^\g(\s_{\m\n})_{\g\b}\H^\b\\
\dmn \chi_{\r\s}&=&\frac{1}{8}\left(\ve_{\m\n\s\tau}+g_{\m\s}g_{\n\tau}-
g_{\m\tau}g_{\n\s}\right)\left[\f,\ov\f\right] \nonumber \\
&&+\left(F_{\m\s}^+ g_{\n\tau}-F_{\n\s}^+ g_{\m\tau}-F_{\m\tau}^+ g_{\n\s}+
F_{\n\tau}^+g_{\m\s}\right) \nonumber \\
&&+\left(\ve _{\m\n\s}^{\ \ \ \ \a}F_{\tau\a}^+ -
\ve_{\m\n\tau}^{\ \ \ \ \a }F_{\s\a}^+ +\ve_{\s\tau\m}^{\ \ \ \ \a}F_{\n\a}^+ 
-\ve_{\s\tau\n}^{\ \ \ \ \a }F_{\m\a}^+\right) \nonumber\\
&&-\frac{1}{16}[g_{\m\s}(\s_{\r\n})_\g^{\ \b}-
g_{\m\r}(\s_{\s\n})_\g^{\ \b}\nonumber\\
&&+g_{\r\n}(\s_{\s\n})_\g^{\ \b}+
g_{\s\n}(\s_{\m\r})_\g^{\ \b}]\left(\Hb_\b\H^\g+\Hb^\g\H_\b\right)\ ,
\nonumber
\end{eqnarray} 
all the other transformations remaining unchanged.

Taking into account the above transformations, now the whole algebra 
formed by the twisted operators $(\dw,\dmu,\dmn)$ is closed, $modulo$ 
field dependent gauge transformations and field equations, for 
the whole theory, including matter.

\subsection{Symmetries of the complete twisted action}

Lengthy and uninstructive computations lead us to claim that the 
complete, twisted action
\bea 
S_{T}=S_{TYM}+S_{Tmatter}\ ,
\label{taction}\eea
where $S_{TYM}$ and $S_{Tmatter}$ are given in \equ{TYM} and 
\equ{matterat} respectively, 
is invariant under the twisted supercharges:
\begin{equation}\label{dwS=0}
\dw S_{T}=\dmu S_{T}=\dmn S_{T}=0\ .
\end{equation}
As for the twisted pure gauge N=2 SYM action, the total $S_{T}$ 
action is univocally determined by the two symmetries $\dw$ and 
$\dmu$, $\dw$ alone being not sufficient. Only one coupling 
constant is left, as expected.

Starting from the untwisted N=2 SYM theory coupled to matter, through 
the twisting procedure we got an action, gauge invariant, and 
invariant as well under two symmetries $\dw$ and $\dmu$, remnant of 
five over the eight supercharges, three of them turning out to be 
redundant. The resulting action is equivalent to the starting, 
untwisted supersymmetric action, and the twist revealed an algebraic 
topological structure.

The twisted N=2 sym action coupled to matter appears to be a 
Witten-type topological action, 
since it can be written as the variation of a nilpotent 
operator ($\dw$, in our case), $modulo$ field equations:

\bea
S_{T} = \dw \Delta + \widehat\Delta\ ,
\label{S=dw(qc)}
\eea
where
\begin{eqnarray}
\Delta &=& 
\tr\int d^{4}x\Big(\ha\chi^{\m\n}F^+_{\m\n} - \ha\fb D^\m\p_\m + 
\frac{1}{16}\eta\left[\f,\fb\right] 
\nonumber\\&&
+ 
\frac{1}{16}\Hb^g(\s_{\m\n})_\g^\b\chi^{\m\n}\H_\b + 
\frac{1}{16}\ub^\g\fb\H_\g + \frac{1}{16} u_\g\fb\Hb^\g\Big)\ ,
\end{eqnarray}
and $\widehat\Delta$ is a contact term
\bea\
\widehat\Delta = 
\ha\chi^{\m\n}\frac{\delta{S}_{T}}{\delta\chi^{\m\n}} + 
v^\dg\frac{\delta{S}_{T}}{\delta v^\dg} + 
\vb^\dg\frac{\delta{S}_{T}}{\delta\vb^\dg}\ .
\eea

We recall that $\dw$ is nilpotent only {\it on shell}, and the above 
relation is not yet an {\it exact} relation, in the cohomological 
sense. But, still, 
this last result is quite remarkable. It suggests indeed to consider the 
operator $\dw$ -- which, we stress again, is not sufficient, alone, to 
completely determine the action -- as the starting point towards the identification
of a nilpotent operator under which the total action is
{\it off shell} exact.

\section{Towards quantum extension}

In the previous sections, we managed in order to treat a known, though complex, 
situation. Namely we are now dealing with the gauge invariant action $S_{T}$ 
\equ{taction}, invariant also under the scalar operator $\dw$ and the vector 
operator $\dmu$. The underlying
supersymmetry algebra closes on translations, $modulo$ field 
equations and gauge dependent field transformations. The situation is 
similar to that encountered in topological field theories (like Chern-Simons 
theory or BF models) and in supersymmetric field theories 
(N=1 and N=2 SYM). We 
shall adopt in this case the same technique successfully used there, 
to define the classical theory and to 
proceed towards the algebraic renormalization.

The study of the divergences of a quantum field theory and of the 
possible quantum extension of its classical symmetries requires the 
usual renormalizations tools. In the case of supersymmetric field 
theories, so far it is not known a completely satisfactory 
regularization scheme which preserves at the same time BRS 
symmetry and supersymmetry. 
The algebraic renormalization~\cite{Piguet:1995er}, which 
does not rely on any regularization scheme, is, hence, a mandatory 
choice. The first algebraic study of the renormalizability of a 
supersymmetric QFT, has been completely performed, for the N=1 case, using the 
superspace formalism (see \cite{Piguet:1986ug} and references therein, 
in particular \cite{Piguet:1984aa}), and a class of N=1 SYM theories has been 
shown to have no coupling constant renormalization at all 
\cite{Piguet:1986pk,Piguet:1986td,Lucchesi:1987ef,Parkes:1984dh,Parkes:1985hj}.

For what 
concerns N=2 SYM, with and without matter, the first algebraic 
approach to the study of counterterms and anomalies
has been given in \cite{Maggiore:1994dw,Maggiore:1994xw}. 

In this Section, we set the standard for the quantum extension of the 
twisted N=2 TYM, all the results obtained previously thanks to the 
twist being valid at the 
classical level only. The basic steps of the procedure are the 
construction of an invariant gauge fixing term, the definition of a 
classical action, including gauge fixing and source-dependent 
terms,  which satisfies all the symmetries of the theory through an 
extended Slavnov-Taylor identity, which resumes both gauge symmetries and 
supersymmetries. The key point in our reasoning is the closure of the 
algebra {\it off shell}.

\subsection{TYM and the extended BRS operator}

Our starting point is the classical  action $S_{T}$ \equ{taction}, 
equivalent to the
classical N=2 SYM coupled to matter.

Besides being gauge invariant, the action $S_{T}$ is invariant also 
under a set of global transformations, whose generators $\dw,\dmu,\dmn$
commute with the gauge transformations $\delta^g_\e$~\equ{g-transf},
and satisfy the following algebra:
\begin{eqnarray}\label{algebra}
\left[\dw,\delta^g_\e\right]&=&\left[\dmu,\delta^g_\e\right]=
\left[\dmn,\delta^g_\e\right]=0\nonumber\\
\dw^2&=&\delta_\f^g\;+\;(\textrm{field eq.})
\nonumber\\
\left\{ \dmu ,\dnu \right\}&=&-\frac{1}{8}g_{\m\n}\delta_\fb^g\;+
(\textrm{field eq.})\nonumber\\
\left\{ \dw,\dmu \right\}&=&\d_\m +\delta_{A_\m}^g+
(\textrm{field eq.})
\\
\left\{\dw,\dmn\right\}&=&(\textrm{gauge transf.})+
(\textrm{field eq.})\nonumber\\
\left\{\dmn,\drs\right\}&=&(\textrm{gauge transf.})+
(\textrm{field eq.})\nonumber\\
\left\{ \dmu ,\drs\right\}&=&-(\ve_{\m\r\s\n}\d ^\n +g_{\m\r}\d_\s -
g_{\m\s}\d_\r)+(\textrm{gauge transf.})+(\textrm{field eq.})\ .
\nonumber
\end{eqnarray}
The quantum extension of susy, or susy-like, theories presents some 
serious difficulties, as explained in \cite{Breitenlohner:1985kh}:
\begin{description}
\item{\bf gauge fixing term:} 
In absence of supersymmetry, the gauge fixing term is
a BRS variation, hence it is BRS invariant by construction, being the 
BRS operator nilpotent. In 
presence of supersymmetry, instead, because of the algebra, which in 
the WZ gauge does not simply close on translations, such a term is 
not susy invariant. The usual way to add a gauge fixing term cannot 
be applied for supersymmetric QFT.
\item{\bf open algebra:} 
In the Wess Zumino gauge, the susy transformations (and hence their 
twisted versions) are not linear. The algebra closes only {\it on shell }
and $modulo$ field dependent gauge transformations. The standard way 
to deal with this kind of algebras is to introduce auxiliary fields 
in order to get rid of the field equations, but still the algebra 
does not close, and an infinite number of external fields is needed, 
which renders the quantum extension of the theory meaningless.
\end{description} 
A solution is to define an extended BRS operator 
which collects all the symmetries of 
the theory~\cite{Maggiore:1994dw,Maggiore:1994xw}, 
but, before doing that, let us write the usual BRS operator $s$, 
promoting the 
gauge parameter $\e(x)$ to a ghost field $c(x)$, so that 
the gauge transformation 
$\delta_\e^g$ becomes the BRS operator $s$: 
\bea
\e^a(x)\rightarrow c^a(x)\ \ ,\ \ \delta^g_\e \rightarrow s\ .
\eea
In addition, we introduce an antighost field $\ov c(x)$ and a 
Lagrange multiplier $b(x)$, always in the adjoint representation of 
the gauge group, so that the BRS operator 
\begin{eqnarray}\label{BRStot}
sA_\mu &=&-D_\mu c \nonumber \\
s\psi _\mu &=&\left\{ c,\psi _\mu \right\}  \nonumber \\
s\chi _{\mu \nu } &=&\left\{ c,\chi _{\mu \nu }\right\}\nonumber \\
s\eta &=&\left\{ c,\eta \right\}\nonumber \\
s\phi &=&\left[ c,\phi \right]   \nonumber \\
s\overline{\phi } &=&\left[ c,\overline{\phi }\right] \nonumber \\
sc &=&c^2=\ha f^{abc}c^b c^c\nonumber\\
s \ov c&=&b \\
s b&=&0\nonumber\\
s\H &=&\left[ c,\H \right] \nonumber \\
s\Hb &=&\left[ c,\Hb \right]  \nonumber \\
s u &=&\left\{ c, u \right\} \nonumber \\
s\ub &=&\left\{ c,\ub \right\}\nonumber \\
s v &=&\left\{ c, v \right\} \nonumber \\
s\vb &=&\left\{ c,\vb \right\} \nonumber 
\end{eqnarray}
is nilpotent
\bea
s^{2}=0\ .
\eea
Let us now introduce $global$ ghosts $\omega$, $\e^{\m}$ and $v^{\m}$, 
coupled respectively to $\dw$, $\dmu$ and to the 
translations $\d_{\m}$
\begin{equation}\label{paramSUSY}
\omega \leftrightarrow \delta _{\mathcal{W}}\;,\;\;\;\varepsilon ^\mu
\leftrightarrow \delta _\mu \;,\;\;\;v^\mu \leftrightarrow \partial 
_\mu \;,\;
\end{equation}
in order to define the $extended$ BRS operator as
\begin{equation}\label{BRSext}
\mathcal{Q=}s+\omega \delta _{\mathcal{W}}+\varepsilon ^\mu \delta _\mu
+v^\mu \partial _\mu -\omega \varepsilon ^\mu \frac \partial {\partial v^\mu
}\ .
\end{equation}
The mass dimensions, ${\cal R}$-charge, ghost number and the statistics 
of the ghosts (both global and local), of the antighost and of the Lagrange multiplier,
are summarized in the following table
\[
\begin{tabular}{|c|c|c|c|c|c|c|}
\hline
&  $\omega $ & $\varepsilon ^\mu $ & $v^\mu$ &c&  $\bar{c}$ & $b$  \\ \hline
$\dim $ & $-1/2$ & $-1/2$ & $-1$ &0 & 2 & 2 \\ \hline
$\mathcal{R-}\textrm{charge}$ & $-1$ & $1$ & $0$ & 0 & 0 & 0  \\ \hline
$\mathrm{ghost\ number}$ & $1$ & $1$ & $1$ &1 & $-1$ & 0 \\ \hline
$\mathrm{statistics}$  & $comm$ & $comm$ & $ant$ &$ant$ & $ant$ & $comm$ \\ 
\hline
\end{tabular}
\]
The extended BRS operator ${\cal Q}$ has ghost number +1, zero ${\cal 
R}$-charge and mass dimensions, is a symmetry of the action $S_{T}$ 
and it is nilpotent $on\ shell$:
\begin{eqnarray}
\mathcal{Q}\; {S}_{T} &=&0\;,  \label{QQ} \\
\mathcal{Q}^2 &=&\mbox{field equations}\;.  \nonumber
\end{eqnarray}
The ${\cal Q}$-invariance of the action is obvious, since $S_{T}$ 
does not depend on the ghosts and it is invariant under translations. 
On the other hand, the nilpotency of ${\cal Q}$ is obtained defining 
the action of the twisted operators $\dw,\dmu$ (and hence of
${\cal Q}$) on the ghosts 
$\left( c(x),\omega,\varepsilon ^\mu ,v^\mu \right)$ suitably. 

It must be
\begin{eqnarray}\label{Qghosts}
\mathcal{Q}c &=&c^2-\omega ^2\phi -\omega \varepsilon ^\mu A_\mu +\frac{%
\varepsilon ^2}{16}\overline{\phi }+v^\mu \partial _\mu c \nonumber\\ 
\mathcal{Q}\omega &=&0\\
\mathcal{Q}\varepsilon ^\mu &=&0\;  \nonumber\\
\mathcal{Q}v^\mu &=&-\omega \varepsilon ^\mu \;.  \nonumber
\end{eqnarray}
The antighost $\ov c(x)$ and the Lagrange multiplier $b(x)$ form a 
${\cal Q}$-doublet:
\begin{eqnarray}
\mathcal{Q}\overline{c}\; &=&b+v^\mu \partial _\mu \overline{c}\;,
\label{cb-b} \\
\mathcal{Q}b &=&\omega \varepsilon ^\mu \partial _\mu \overline{c}+v^\mu
\partial _\mu b\;,  \nonumber
\end{eqnarray}
with
\begin{equation}
\mathcal{Q}^2\overline{c}=\mathcal{Q}^2b=0\;.  \label{Q-cb-b}
\end{equation}

At this point we are able to define a gauge fixing term, as the ${\cal 
Q}$-variation of the usual ``gauge fermion'':
\begin{eqnarray} \label{landau-g-fi}
S_{gf} &=&Q\ \Tr \int d^4x\;\overline{c}\partial A  \\
&=&\Tr\int d^4x\;\left( b\partial ^\mu A_\mu +\overline{c}\partial ^\mu D_\mu
c-\omega \overline{c}\partial ^\mu \psi _\mu -\frac{\varepsilon ^\nu }2%
\overline{c}\partial ^\mu \chi _{\nu \mu }-\frac{\varepsilon ^\mu }8%
\overline{c}\partial _\mu \eta \right) \;.  \nonumber
\end{eqnarray}
Since the extended BRS operator $\cal Q$ is strictly nilpotent on the 
fields appearing in $S_{gf}$, the gauge fixed action $\mathcal{S}$ is ${\cal Q}$-invariant by construction:  
\begin{equation}
\mathcal{Q}\left( \mathcal{S}\right) =
\mathcal{Q}\left( {S}_{T}+S_{gf}\right) =0\;.  \label{Q-g-f-action}
\end{equation}
The gauge fixing procedure takes into account not only the local pure 
gauge symmetry, but also the twisted symmetries $\dw$ and $\dmu$, as 
can be seen by the presence in the gauge fixing term \equ{landau-g-fi} of the global 
ghosts $\omega$ and $\e^{\m}$. The absence of $v^{\m}$ is due to the 
translation invariance.

Therefore, the action of the extended BRS operator ${\cal Q}$ on the whole set 
of fields and ghosts, is: 
\begin{eqnarray}
\mathcal{Q}A_\mu &=&-D_\mu c+\omega \psi _\mu +\frac{\varepsilon ^\nu }2\chi
_{\nu \mu }+\frac{\varepsilon _\mu }8\eta +v^\nu \partial _\nu A_\mu 
\nonumber\\
\mathcal{Q}\psi _\mu &=&\left\{ c,\psi _\mu \right\} -\omega D_\mu \phi
+\varepsilon ^\nu \left( F_{\nu \mu }-\frac 12F_{\nu \mu }^{+}\right) -
\frac{\varepsilon _\mu }{16}[\phi ,\overline{\phi }]  \nonumber \\
&&+v^\nu \partial _\nu \psi _\mu +
\frac{1}{16}\Hb^\g(\s_{\n\m})_{\g\b}\H^\b\ve^\n\nonumber\\
\mathcal{Q}\chi _{\sigma \tau } &=&\left\{ c,\chi _{\sigma \tau }\right\}
+\omega F_{\sigma \tau }^{+}+\frac{\varepsilon ^\mu }8(\varepsilon _{\mu
\sigma \tau \nu }+g_{\mu \sigma }g_{\nu \tau }-g_{\mu \tau }g_{\nu \sigma
})D^\nu \overline{\phi }\;  \nonumber \\
&&+v^\nu \partial _\nu \chi _{\sigma \tau }-
\frac{\omega}{8}\Hb^\g(\s_{\s\tau})_{\g\b}\H^\b\nonumber\\
\mathcal{Q}\eta &=&\left\{ c,\eta \right\} +
\frac \omega 2[\phi ,\overline{%
\phi }]+\frac{\varepsilon ^\mu }2D_\mu \overline{\phi }+
v^\nu \partial _\nu
\eta   \nonumber \\
\mathcal{Q}\phi &=&\left[ c,\phi \right] -\varepsilon ^\mu \psi _\mu +
v^\nu
\partial _\nu \phi   \nonumber \\
\mathcal{Q}\overline{\phi } &=&\left[ c,\overline{\phi }\right] +2\omega
\eta +v^\nu \partial _\nu \overline{\phi }  \nonumber \\
\mathcal{Q}c &=&c^2-\omega ^2\phi -\omega \varepsilon ^\mu A_\mu +\frac{%
\varepsilon ^2}{16}\overline{\phi }+v^\nu \partial _\nu c  \nonumber \\
\mathcal{Q}\omega &=&0\;\;\;  \nonumber \\
\mathcal{Q}\varepsilon ^\mu &=&0\;\;\;  \label{Q-transf} \\
\mathcal{Q}v^\mu &=&-\omega \varepsilon ^\mu   \nonumber \\
\mathcal{Q}\overline{c}\; &=&b+v^\mu \partial _\mu \overline{c}\; 
\nonumber \\
\mathcal{Q}b &=&\omega \varepsilon ^\mu \partial _\mu \overline{c}+v^\mu
\partial _\mu b  \nonumber\\
{\cal Q}\H_\g&=&\left[c,\H_\g\right]+\omega u_\g+\ve^\m(\s_\m)_{\g\da}v^\da+
v^\m\d_\m\H_\g\nonumber\\
{\cal Q}\Hb_\g&=&\left[c,\Hb_\g\right]+\omega \ub_\g+
\ve^\m(\s_\m)_{\g\da}\vb^\da+v^\m\d_\m\Hb_\g\nonumber\\
{\cal Q}u_\g&=&\left[c,u_\g\right]+\omega\f\H_\g+
\ve^\m\left(\ha D_\m\H_\g-\ha(\s_{\m\n})_\g^\b D^\n\H_\b\right)+
v^\m\d_\m u_\g\nonumber\\
{\cal Q}\ub_\g&=&\left[c,\ub_\g\right]+\omega\f\Hb_\g+
\ve^\m\left(\ha D_\m\Hb_\g-\ha(\s_{\m\n})_\g^\b D^\n\Hb_\b\right)+
v^\m\d_\m \ub_\g\nonumber\\
{\cal Q}v_\dg&=&\left[c,v_\dg\right]-\ha\omega (\s^\n)_{\a\dg}D_\n\H^\a-
\frac{1}{16}\ve^\m(\sb_\m)_{\dg\b}\fb\H^\b+v^\m\d_\m v_\dg\nonumber\\
{\cal Q}\vb_\dg&=&\left[c,\vb_\dg\right]-
\ha\omega (\s^\n)_{\a\dg}D_\n\Hb^\a-
\frac{1}{16}\ve^\m(\sb_\m)_{\dg\b}\fb\Hb^\b+v^\m\d_\m\vb_\dg\ , \nonumber
\end{eqnarray}
%
with
\begin{equation}
\mathcal{Q}^2=0\;\;\;\;\mathrm{on\;\;\;\;}\left( A,\phi ,\overline{\phi }%
,\eta ,\H,\Hb,c,\omega ,\varepsilon ,v,\overline{c},b\right) \;,  \label{QQ-1}
\end{equation}
and
\begin{eqnarray}
\mathcal{Q}^2\psi _\sigma &=&\frac{g^2}4\omega \varepsilon ^\mu \frac{\delta 
\mathcal{S}}{\delta \chi ^{\mu \sigma }}  \label{QQ-2} \\
&&+\frac{g^2}{32}\varepsilon ^\mu \varepsilon ^\nu \left( g_{\mu \sigma }%
\frac{\delta \mathcal{S}}{\delta \psi ^\nu }+g_{\nu \sigma }\frac{%
\delta \mathcal{S}}{\delta \psi ^\mu }-2g_{\mu \nu }\frac{\delta 
\mathcal{S}}{\delta \psi ^\sigma }\right)  \nonumber
\end{eqnarray}
\begin{eqnarray}
\mathcal{Q}^2\chi _{\sigma \tau } &=&-\frac{g^2}2\omega ^2\frac{\delta 
\mathcal{S}}{\delta \chi ^{\sigma \tau }}  \label{QQ-3} \\
&&+\frac{g^2}8\omega \varepsilon ^\mu \left( \varepsilon _{\mu \sigma \tau
\nu }\frac{\delta \mathcal{S}}{\delta \psi _\nu }+g_{\mu \sigma }\frac{%
\delta \mathcal{S}}{\delta \psi ^\tau }-g_{\mu \tau }\frac{\delta 
\mathcal{S}}{\delta \psi ^\sigma }\right)  \nonumber
\end{eqnarray}
\begin{eqnarray}\label{QQu}
{\cal Q}^2 u_\g&=&
\frac{g^2}{2}\Big(\omega\ve^\m(\s_\m)_{\g\dg}\frac{\delta \mathcal{S}}{\delta \vb^\dg} + 
\ve^2\frac{\delta \mathcal{S}}{\delta \ub^\g}\Big)
\end{eqnarray}
\begin{eqnarray}\label{QQub}
{\cal Q}^2 \ub_\g&=&
\frac{g^2}{2}\Big(\omega\ve^\m(\s_\m)_{\g\dg}\frac{\delta \mathcal{S}}{\delta v^\dg} + 
\ve^2\frac{\delta \mathcal{S}}{\delta u^\g}\Big)
\end{eqnarray}
\begin{eqnarray}\label{QQv}
{\cal Q}^2 v_\dg&=&\frac{g^2}{2}\Big(\omega^2\frac{\delta \mathcal{S}}{\delta \vb_\dg} - 
\omega\ve^\m(\s_\m)_{\b\dg}\frac{\delta \mathcal{S}}{\delta \ub_\b}\Big)
\end{eqnarray}
\begin{eqnarray}\label{QQvb}
{\cal Q}^2 \vb_\dg&=&\frac{g^2}{2}\Big(\omega^2\frac{\delta \mathcal{S}}{\delta v_\dg} - 
\omega\ve^\m(\s_\m)_{\b\dg}\frac{\delta \mathcal{S}}{\delta u_\b}\Big)\ .
\end{eqnarray}

\subsection{The Slavnov-Taylor identity}

For the functional implementation of the extended BRS operator 
${\cal Q}$, we must couple external sources $\Phi^{\star i}(x)$ to the nonlinear ${\cal 
Q}$-transformations of the fields $\Phi^{i}(x)$ \equ{Q-transf}:
\[
\begin{array}{ccc}
L\rightarrow c & \quad , \quad  &  X^\g\rightarrow\H_\g \\
D\rightarrow \f & \quad , \quad & \ov X^\g\rightarrow\Hb_\dg \\
\Omega^\m\rightarrow A_\m & \quad , \quad & U^\g\rightarrow u_\g \\
\xi^\m\rightarrow \p_\m & \quad , \quad & \ov U^\g\rightarrow\ub_\g \\
\rho\rightarrow \fb & \quad , \quad & V^\dg\rightarrow v_\dg \\
\tau\rightarrow \eta & \quad , \quad & \ov V^\dg\rightarrow\vb_\dg \\
B^{\m\n}\rightarrow\chi_{\m\n}&&\ ,
\end{array}
\]
so that we can add to ${\cal S} = S_{T} + S_{gf}$ the ``external'' term
\bea
S_{ext} = 
\Tr\int d^4x\; 
\Phi^{\star i}\mathcal{Q}\Phi^{i}\ ,
\label{S-sorg}\eea
where we collectively denoted with $\Phi^{i}(x)$ all the fields transforming 
nonlinearly under ${\cal Q}$, and with $\Phi^{\star i}(x)$ the corresponding 
external sources, whose quantum numbers and statistics are
\[
\begin{tabular}{|c|c|c|c|c|c|c|c|}
\hline
& $L$ & $D$ & $\Omega ^\mu $ & $\xi ^\mu $ & $\rho $ & $\tau $ & $B^{\mu \nu
}$ \\ \hline
$\dim .$ & $4$ & $3$ & $3$ & $5/2$ & $3$ & $5/2$ & $5/2$ \\ \hline
$\mathcal{R}-\textrm{charge}$ & $0$ & $-2$ & $0$ & $-1$ & $2$ & $%
1 $ & $1$ \\ \hline
$\mathrm{ghost number}$ & $-2$ & $-1$ & $-1$ & $-1$ & $-1$ & $-1$ & $-1$ \\ 
\hline
$\mathrm{statistics}$ & $comm$ & $ant$ & $ant$ & $comm$ & $ant$ & $comm$ & 
$comm$ \\ \hline
\end{tabular} 
\]
\[
\begin{tabular}{|c|c|c|c|c|c|c|}
\hline
& $X$ & $\ov X$ & $U$ & $\ov U$ & $V$ & $\ov V$ \\ \hline
$\dim .$ & $3$ & $3$ & $5/2$ & $5/2$ & $5/2$ & $5/2$ \\ \hline
$\mathcal{R}-\textrm{charge}$ & $0$ & $0$ & $-1$ & $-1$ & $1$ & $1$ \\ \hline
$\mathrm{ghost number}$ & $-1$ & $-1$ & $-1$ & $-1$ & $-1$ & $-1$ \\ \hline
$\mathrm{statistics}$ & $ant$ & $ant$ & $comm$ & $comm$ & $comm$ & $comm$ \\ 
\hline
\end{tabular} 
\]
In order to write a Slavnov-Taylor identity, the last step is to add 
to the action $S_{T} + S_{gf} + S_{ext}$ a fourth term $S_{quad}$ 
which takes 
into account the fact the the extended BRS operator ${\cal Q}$ is 
nilpotent $on\ shell$, according to the Batalin-Vilkokisky procedure 
\cite{Batalin:1981jr,Batalin:1984jr}. 
Such a term must be quadratic in the external sources 
$\Phi^{*i}(x)$
\begin{equation}\label{Squad}
S_{quad}={\rm Tr}\xint\left(\Omega_{ij}\Phi^{*i}\Phi^{*j}\right) \ ,
\end{equation}
where $\Omega_{ij}(x)$ are coefficients which, in general, may depend on 
the quantum fields and on the global ghosts. They are determined by 
imposing the validity of the Slavnov-Taylor identity, which we shall 
write shortly. 

The result is the following: 
\begin{eqnarray}\label{s-quad}
S_{quad}&=&g^{2}\Tr\int d^4x\Big( \frac 1 8\omega ^2B^{\mu \nu
}B_{\mu \nu }-\frac 14\omega B^{\mu \nu }\varepsilon _\mu \xi _\nu
-\frac  1 {32}\varepsilon ^\mu \varepsilon ^\nu \xi _\mu \xi _\nu +\frac
1{32}\varepsilon ^2\xi ^2   \nonumber\\
&&-\frac{1}{2}\ve^2 U_\g \ov U^\g + \frac{1}{2}\omega^2 V_\dg \ov V^\dg - 
\frac{1}{2}\omega\ve^\m \ov U^\a(\s_\m)_{\a\dg}V^\dg \Big)\ .
\end{eqnarray}
With this choice, the complete classical action
\begin{equation}
\Sigma =S_{T} + S_{gf} + S_{ext} +  S_{quad}
\label{c-action}
\end{equation}
satisfies the Slavnov-Taylor identity
\begin{equation}
\mathcal{S}(\Sigma )\;\mathcal{=\;}0\;,  \label{tym-s-t}
\end{equation}
where
\begin{eqnarray}
{\cal S}(\Sigma ) &=& \Tr\int d^4x\left(
\frac{\delta \Sigma }{\delta \Phi^{\star i} 
}\frac{\delta \Sigma }{\delta \Phi^{i}}
+(b+v^\mu \partial_\mu \overline{c})\frac{\delta \Sigma }{\delta \overline{c}} 
 +(\omega \varepsilon ^\mu \partial _\mu 
\overline{c}+v^\mu \partial _\mu b)\frac{\delta \Sigma }{\delta b}\right)
\nonumber \\
&&-\omega \varepsilon ^\mu \frac{\partial \Sigma }{\partial v^\mu }\;.
\label{tym-s-t-op}
\end{eqnarray}
We can go further, introducing the translation operator
\bea
{\cal P}_\mu \Sigma = \tr
\int d^4x\left( \partial _\mu
\Phi ^i\frac{\delta \Sigma }{\delta \Phi ^i}
+\partial _\mu \Phi^{*i}\frac{\delta \Sigma }{\delta \Phi ^{*i}}\right) =\;0\ ,
\label{transl-inv}
\eea
which obviously is a symmetry of the theory.

We observe that, since ${\cal P}_{\m}$ acts linearly on all the fields, 
the dependence of the action $\S$ on the global ghost for 
translations $v_{\m}$, is fixed by the following identity
\begin{equation}
\frac{\partial \Sigma }{\partial v^\mu }=\Delta _\mu ^{cl}\;,  \label{v-id}
\end{equation}
where
\begin{eqnarray}
\Delta _\mu ^{cl} &=&\tr\int d^4x(\;L\partial _\mu c-D\partial _\mu \phi
-\Omega ^\nu \partial _\mu A_\nu +\xi ^\nu \partial _\mu \psi _\nu \; 
\nonumber \\
&&-\rho \partial _\mu \overline{\phi }+\tau \partial _\mu \eta + 
B^{\nu \sigma }\partial _\mu \chi _{\nu \sigma} - X^\g\d_\m \H_\g - 
\ov X^\g\d_\m \Hb_\g \nonumber\\
&&+ U^\g\d_\m u_\g + \ov U^\g\d_\m \ub_\g + V^\dg\d_\m v_\dg + 
\ov V^\dg\d_\m \vb_\dg \;)\ , \label{v-break}
\end{eqnarray}
being linear in the quantum fields, is present only at the classical 
level~\cite{Piguet:1995er}. 
We can 
therefore get rid of the global ghost $v_{\m}$, introducing the 
``reduced'' classical action $\widehat{\Sigma }$ 
\bea
\Sigma =\widehat{\Sigma }+v^\mu \Delta _\mu ^{cl}\;,  \label{h-action} 
\eea
with
\bea
\frac{\partial \widehat{\Sigma }}{\partial v^\mu } =0\;.  
\eea
It is easily verified that $\widehat{\Sigma }$ satisfies the 
modified ST identity
\begin{equation}
\mathcal{S}(\widehat{\Sigma })\;\mathcal{=\;}\omega \varepsilon ^\mu \Delta
_\mu ^{cl}\;\;,  \label{n-tym-s-t}
\end{equation}
where
\bea
{\cal S}(\widehat{\Sigma } ) = \Tr\int d^4x\left(
\frac{\delta \widehat{\Sigma }}{\delta \Phi^{\star i} 
}\frac{\delta \widehat{\Sigma } }{\delta \Phi^{i}}
+b\frac{\delta \widehat{\Sigma }}{\delta \overline{c}} 
 +\omega \varepsilon ^\mu \partial _\mu 
\overline{c}\frac{\delta \widehat{\Sigma } }{\delta b}\right)\;.
\label{n-tym-st-op}
\end{eqnarray}

The (classically broken) ST identity \equ{n-tym-s-t} is the one which 
must be  
used to determine the quantum extension of the theory. The corresponding 
linearized ST operator
\bea
\mathcal{B}_{\widehat{\Sigma }} =
\Tr\int d^4x\left(
\frac{\delta \widehat{\Sigma }}{\delta \Phi^{i}}
\frac \delta {\delta \Phi^{\star i}}
+
\frac{\delta \widehat{\Sigma }}{\delta \Phi^{\star i} }
\frac \delta {\delta \Phi^{i}}
+
b\frac{\delta \widehat{\Sigma }}{\delta \overline{c}} 
 +\omega \varepsilon ^\mu \partial _\mu 
\overline{c}\frac{\delta \widehat{\Sigma } }{\delta b}\right)\ ,
\label{n-tym-lin-op}
\eea
is not nilpotent. In fact, it holds
\begin{equation}
\mathcal{B}_{\widehat{\Sigma }}\mathcal{B}_{\widehat{\Sigma }}=\omega
\varepsilon ^\mu \mathcal{P}_\mu \;,  \label{nil-lin-tym}
\end{equation}
that is, $\mathcal{B}_{\widehat{\Sigma }}\;$ is nilpotent $modulo$ a 
total derivative. It follows that the operator 
$\mathcal{B}_{\widehat{\Sigma }}\;$ is nilpotent in the
space of $integrated$ local functionals, which, actually, 
is the case we are interested in.

Summarizing, we handled the problem in order to be able to deal
with the usual web of symmetries and constraints 
which constitutes the 
basis for the quantum extension of the model and for the study of its 
algebraic renormalizability (determination of local counterterms 
and study of anomalies)~\cite{Piguet:1995er}: 

\begin{itemize}
\item \textbf{ST identity \equ{n-tym-s-t}}\ ;
\item \textbf{Landau gauge fixing condition}
\begin{equation}
\frac{\delta \widehat{\Sigma }}{\delta b}=\d^\m A_\m\;; 
\label{landau-g-f}
\end{equation}
\item \textbf{Anti-ghost equation}
\begin{equation}
\frac{\delta \widehat{\Sigma }}{\delta \overline{c}}+\partial _\mu \frac{%
\delta \widehat{\Sigma }}{\delta \Omega _\mu }=0\;;  \label{antigh-eq}
\end{equation}
\item \textbf{Landau gauge ghost equation}
\begin{equation}
\tr\int d^4x\left( \frac{\delta \widehat{\Sigma }}{\delta c}+\left[ \overline{c}%
,\frac{\delta \widehat{\Sigma }}{\delta b}\right] \right) =\Delta _c^{cl}\;,
\label{gh-w-id}
\end{equation}
with $\Delta _c^{cl}\;$ linear classical breaking
\begin{eqnarray}
 \Delta _c^{cl}&=&\tr\int d^4x\left( [c,L]-[A,\Omega ]-[\phi ,D]+[\psi ,\xi ]-
 [\overline{\phi },\rho ]+[\eta ,\tau ] \right.\\
 && \left. +[\chi ,B] - [\H,X] - [\Hb,\ov X] + [u,U] + [\ub, \ov U] + [v,V] 
 + [\vb,\ov V]\right)\ .
 \nonumber
 \label{gh-lin-break}
 \end{eqnarray}
\end{itemize}

\section{Conclusions}

In this paper we studied the twisted version of N=2 SYM theory coupled to matter. 

The twist, being simply a linear redefinition of the quantum fields, does not affect 
the partition function, and hence two 
twisted-related theories
are completely equivalent. This means, in particular, that they have the same physical 
content, the same observables and the same coupling constant(s) $\beta$-function(s).

In this paper we included matter into the game. We twisted the hypermultiplet, which 
became entirely spinorial, we modified the (twisted) supersymmetries in order to have 
a close off-shell algebra, and we achieved the complete off-shell set up by means of 
a unique Slavnov-Taylor identity, which collects both BRS symmetry and supersymmetries 
of the theory.

As it is well known \cite{Witten:1988ze,Witten:1994ev}, pure N=2 SYM is twisted to a 
topological quantum field theory: TYM. An interesting and new result presented in this 
article is the fact that TYM theory coupled to matter have the same set of invariances 
of the same theory without matter. Since the theory with matter is not topological, the 
presence of these symmetries contradicts the common belief that they are peculiar to 
topological theories.\footnote{We thank one of the referees for this remark.} 

The twisted version of the whole theory, including matter, 
is the necessary step towards the study of the $\beta$-function, for which 
a well known non-renormalization theorem holds, and which has been algebraically
proved only in absence of matter \cite{Blasi:2000qw}. We stress also that we never 
specified to which representation of the gauge group the matter hypermultiplet belongs: 
in the particular case of matter  in the adjoint representation, N=4 SYM is recovered.

In general, the introduction of matter spoils the topological character of the theory. 
In our case, the relation \equ{S=dw(qc)} suggests that matter might enter in the theory 
simply through an extended  BRS variation, and this result strongly induces to suppose that matter 
does not alter neither the physical sector of observables nor the finite, or protected, 
operators of the theory \cite{Maggiore:2001zw}. 
In other terms, the presence of matter should not 
spoil the AdS/CFT duality between non-conformal N=2 theories and string theories 
\cite{Polchinski:2000mx}. It is also natural to expect that  the whole action can be written, 
as in the pure gauge case, in terms of a unique, and probably finite, operator, which in the 
pure gauge case is $\tr \phi^{2}$, whose relevance for the algebraic proof of the non-renormalization 
theorem of the $\beta$-function  has been discussed in \cite{Blasi:2000qw}.

\end{document}